\documentclass[preprint,12pt]{elsarticle}
\usepackage[margin=2.5cm]{geometry}
\usepackage{amsmath,amssymb,mathtools}
\usepackage{bm}
\usepackage{booktabs}
\usepackage{array}
\usepackage{xcolor}
\usepackage{graphicx}

\usepackage{pgfplots}
\pgfplotsset{compat=1.17}
\usepgfplotslibrary{groupplots}
\usetikzlibrary{calc}
\usepackage[colorlinks,linkcolor=blue!50!black,citecolor=blue!50!black,urlcolor=blue!50!black]{hyperref}
\pgfplotsset{colormap={rdbu}{rgb255=(5,48,97) rgb255=(33,102,172) rgb255=(67,147,195) rgb255=(146,197,222) rgb255=(247,247,247) rgb255=(244,165,130) rgb255=(214,96,77) rgb255=(178,24,43) rgb255=(103,0,31)}}
\pgfplotsset{colormap={viridis}{rgb255=(68,1,84) rgb255=(72,40,120) rgb255=(62,73,137) rgb255=(49,104,142) rgb255=(38,130,142) rgb255=(31,158,137) rgb255=(53,183,121) rgb255=(110,206,88) rgb255=(181,222,43) rgb255=(253,231,37)}}
\pgfplotsset{colormap={ylorrd}{rgb255=(255,255,255) rgb255=(255,255,204) rgb255=(255,237,160) rgb255=(254,217,118) rgb255=(254,178,76) rgb255=(253,141,60) rgb255=(252,78,42) rgb255=(227,26,28) rgb255=(189,0,38) rgb255=(128,0,38)}}

\definecolor{cRR}{HTML}{0072B2}\definecolor{cCM}{HTML}{D55E00}\definecolor{cRM}{HTML}{009E73}\definecolor{cBGK}{HTML}{555555}
\definecolor{cRRs}{HTML}{CC79A7}\definecolor{cPR}{HTML}{E69F00}\definecolor{cRRL}{HTML}{56B4E9}\definecolor{cCHM}{HTML}{8A2BE2}
\definecolor{cN32}{HTML}{E69F00}\definecolor{cN64}{HTML}{009E73}\definecolor{cN128}{HTML}{0072B2}

\newcommand{\cs}{c_s^2}
\newcommand{\csm}{c_{s,\mathrm m}^2}
\newcommand{\He}{\mathrm{He}}
\newcommand{\bu}{\mathbf u}
\newcommand{\bb}{\mathbf b}
\newcommand{\bc}{\mathbf c}

\newcommand{\ee}{\boldsymbol\epsilon}
\newcommand{\NN}{\mathcal N_D}

\newcommand{\RecA}{\mathcal A}
\newcommand{\nab}{n^{\alpha\beta}}
\newcommand{\eq}{\mathrm{eq}}

\journal{Journal of Computational Physics}

\begin{document}
\begin{frontmatter}

\title{Recursive regularisation in lattice Boltzmann magnetohydrodynamics}

\author{Alessandro De Rosis}
\ead{alessandro.derosis@manchester.ac.uk}
\address{Department of Mechanical and Aerospace Engineering, The University of Manchester, Manchester M13 9PL, United Kingdom}

\begin{abstract}
Recursive regularisation (RR) improves the stability of lattice Boltzmann schemes, but its structure and limitations in magnetohydrodynamics (MHD) remain insufficiently understood. This work connects RR to central-Hermite-moment (CHM) collisions on D2Q9 and D3Q27 through a closed-form reconstruction valid for arbitrary second-order relaxation rates. The central moments of the Maxwell stress coincide, up to sign, with the recursion coefficients, providing a systematic basis for MHD collision operators. Common two-dimensional raw- and central-moment schemes thus correspond to regularisation with a unit bulk relaxation rate. In three dimensions, a CHM implementation reproduces RR to round-off, while a monomial central-moment collision differs at fourth order and produces a four- to twelve-fold larger kinetic-energy error in the tested flow. Chapman--Enskog analysis shows that, with a magnetic field, the hydrodynamic recursion reconstructs the third-order non-equilibrium with a relative error of $78$--$96\%$ ($4\%$ without a field). A local correction largely removes it, yet changes the solutions by less than their discretisation error, consistent with its Burnett-order character. Three-dimensional vortex simulations show approximately second-order convergence at $\mathrm{Re}=100$ and agree with pseudo-spectral solutions through the peak current at $\mathrm{Re}=1000$ and, with a unit bulk rate, at $\mathrm{Re}=3000$. There, single-rate RR develops a compressive instability whose onset depends on the grid, lattice Mach number and initial magnetic configuration. For the standard initial field, a unit bulk rate suppresses it at $\mathrm{Ma}=0.034$ for $N\ge128$. This improvement does not extend to BGK collisions or prevent failure at $\mathrm{Re}=10,000$. These results show that robust three-dimensional MHD simulations require regularisation and bulk dissipation to be considered together.
\end{abstract}

\end{frontmatter}

\section{Introduction}\label{sec:intro}

Magnetohydrodynamics (MHD) describes the dynamics of electrically
conducting fluids interacting with magnetic fields~\cite{biskamp2003mhd}. It provides the
theoretical framework for a wide range of phenomena, from liquid-metal
flows and laboratory plasma experiments to astrophysical and
geophysical processes. In these systems, the fluid motion induces and
transports magnetic fields, while magnetic stresses act back on the
fluid through the Lorentz force. This two-way coupling gives rise to
rich multiscale dynamics, including magnetic-field amplification,
magnetic reconnection, turbulence and the formation of thin current
sheets. Their numerical prediction is particularly challenging at
high Reynolds numbers, where the coexistence of widely separated
spatial and temporal scales demands accurate and robust computational
methods~\cite{davidson2001}.

Numerical simulations of MHD are commonly based on discretisations of
the coupled momentum and induction equations, which read as follows
\begin{align}
\partial_t \mathbf{u} + \mathbf{u}\cdot\nabla \mathbf{u}
&= -\frac{\nabla p}{\rho_0}
+ \nu \nabla^2 \mathbf{u}
+ \frac{(\nabla\times\mathbf{b})\times\mathbf{b}}{\rho_0}, \label{eq:NS_MHD_intro} \\
\partial_t \mathbf{b} + \mathbf{u}\cdot\nabla \mathbf{b}
&= \mathbf{b}\cdot\nabla \mathbf{u}
+ \eta \nabla^2 \mathbf{b}, \label{eq:induction_intro}
\end{align}
supplemented by the solenoidal constraints
\begin{equation}
\nabla\cdot\mathbf{u} = 0,
\qquad
\nabla\cdot\mathbf{b} = 0.
\label{eq:divfree_intro}
\end{equation}
Here $\partial_t$ represents the partial derivative with respect to time $t$, $\nabla$ is the gradient operator, $\mathbf{u}$ and $\mathbf{b}$ denote the velocity and magnetic fields, respectively, $p$ is the hydrodynamic pressure, $\rho_0$ the reference fluid mass density, $\nu$ the kinematic viscosity and $\eta$ the magnetic diffusivity; $\mathbf{b}$ is the magnetic field divided by $\sqrt{\mu_0}$, with $\mu_0$ the vacuum permeability, so that the Lorentz force per unit volume is $(\nabla\times\mathbf{b})\times\mathbf{b}$. Relevant governing parameters are the Reynolds number, the magnetic Reynolds number and the Prandtl number, which are defined as
\begin{equation}
    \mathrm{Re} = \frac{U_0 L_0}{\nu}, \quad \mathrm{Re_m}= \frac{U_0 L_0}{\eta}, \quad \mathrm{Pr_m}= \frac{\nu}{\eta} \equiv \frac{\mathrm{Re}}{\mathrm{Re_m}},
\end{equation}
respectively, where $U_0$ and $L_0$ denote the characteristic velocity and length of the considered problem. Finite-difference, finite-volume and finite-element methods, alongside spectral
approaches, provide established frameworks for solving these
equations. Their implementation must nevertheless address several
interconnected challenges: accurately capturing the nonlinear
interaction between velocity and magnetic fields, controlling
numerical dissipation, and maintaining the solenoidal constraint
$\nabla\cdot\bb=0$. These requirements become particularly demanding
when thin current sheets and small-scale magnetic structures develop,
as numerical stability and resolution can become decisive factors in
the reliability of the computed solution~\cite{mininni2006, FANG2025109400}.

The lattice Boltzmann method (LBM) offers an alternative
computational framework for MHD. Rather than discretising the
macroscopic equations directly, LBM evolves discrete distribution
functions whose moments recover the hydrodynamic fields and their
governing equations~\cite{kruger2016lattice}. Its local collision and streaming operations,
together with its natural parallel structure, make it attractive for
large-scale simulations on modern high-performance computing
architectures. In MHD, fluid and magnetic fields can be represented
through two coupled sets of distribution functions. In Dellar's
double-distribution scheme~\cite{dellar2002}, the magnetic field is
the zeroth moment of vector-valued populations whose equilibrium
contains the induction flux, while the Lorentz force enters the fluid
through the Maxwell stress in the second moment of the fluid
equilibrium. It has since been employed in simulations of MHD turbulence,
Hall-MHD and liquid-metal flows~\cite{pattison2008,breyiannis2004,vahala2008,dellar2013}.

Despite these advantages, the stability of LBM remains a central
concern in low-viscosity regimes. The standard single-relaxation-time
Bhatnagar-Gross-Krook (BGK) collision~\cite{bgk1954} can become unstable as the viscosity
decreases, precisely where resolving the thin current sheets
characteristic of MHD flows becomes essential. This has motivated the
development of more robust collision operators, including
multiple-relaxation-time schemes~\cite{lallemand2000,dhumieres2002},
central moments~\cite{geier2006}, cumulants~\cite{geier2015}, entropic collisions~\cite{karlin2014} and
regularisation techniques. Projective regularisation~\cite{latt2006,zhang2006} retains only the second-order Hermite
contribution to the non-equilibrium distribution. Recursive
regularisation (RR)~\cite{malaspinas2015,coreixas2017} extends this
idea by reconstructing the higher-order non-equilibrium moments from
the second-order ones through recursive relations derived from the
Chapman--Enskog expansion. RR has consequently become an established
approach, including in large-eddy simulations~\cite{jacob2018}.
These collision models have been compared for hydrodynamic flows, theoretically and through their dissipation and linear stability properties~\cite{wissocq2019,coreixas2020,wissocq2020}.
For MHD, central-moment schemes have been developed in two and three
dimensions~\cite{derosis2017,derosis2018,derosis2019,derosis2020,
derosis2021}, whereas RR has only recently been applied to filter
non-hydrodynamic modes in a double-distribution model~\cite{xiong2025}.

The distinction between these collision strategies is less clear-cut than their formulations suggest. In hydrodynamics, Coreixas \textit{et al.}~\cite{coreixas2019} showed analytically, for the isothermal equilibrium on D2Q9 and D3Q27, that single-rate RR is equivalent to projective regularisation in the comoving frame: in moment space, it retains the second-order non-equilibrium while relaxing all higher-order central Hermite moments to equilibrium. They also showed that a cascaded central-moment scheme with unit bulk and higher-order relaxation rates behaves as RR with a unit bulk relaxation rate, and stressed that the stability gained from central-moment collisions must be decoupled from that provided by bulk viscosity. Related interpretations have connected high-order regularisation, cascaded schemes and complete equilibria on standard lattices~\cite{mattila2017,asinari2008,derosis2019luo}. These results suggest that apparently different collision operators may implement closely related filtering mechanisms. However, their implications for MHD have not been established. In particular, the Maxwell stress contributes to the equilibrium moments in a way that depends on the choice of moment frame; the hydrodynamic recursion does not explicitly account for Lorentz coupling; and the relative contributions of regularisation and bulk dissipation to three-dimensional stability remain unclear.

This work addresses these issues through a theoretical and numerical investigation of RR on D2Q9 and D3Q27 lattices. Three questions guide the analysis. First, what is the precise relationship between RR and central-Hermite-moment (CHM) collisions, and how does the Maxwell stress enter the corresponding moment hierarchy? Second, what non-equilibrium contributions are missed when the hydrodynamic recursion is applied to MHD, and do these omissions affect the macroscopic solution? Third, which mechanisms actually determine the robustness of RR in three-dimensional MHD, and where does the method fail?

The main contributions are:

\begin{itemize}
    \item \textit{A closed-form connection between RR and central moments.}
    On D2Q9 and D3Q27, the complete recursive reconstruction is reduced to a compact expression involving only the second velocity derivatives of the product-form equilibrium. This establishes its equivalence to a CHM collision for any reference equilibrium used consistently by both operators, provided that it has the prescribed density and momentum, and with the same second-order relaxation operator, including a separate bulk relaxation rate. The result also identifies, up to a sign, the central moments of the Maxwell stress with the coefficients of the recursive reconstruction. This provides a direct prescription for constructing consistent MHD central-moment equilibria. It further shows that commonly used two-dimensional raw- and central-moment schemes, when their trace is relaxed to equilibrium, are equivalent to projective and recursive regularisation with a unit bulk relaxation rate, respectively. On D3Q27, the analysis isolates the fourth-order discrepancy introduced by the monomial central-moment basis.

    \item \textit{A quantitative assessment of the Lorentz-force contribution.}
    A Chapman--Enskog analysis identifies the third-order non-equilibrium contribution generated by the Maxwell stress and quantifies the fraction missed by the standard hydrodynamic recursion. A local correction is derived to recover this contribution without requiring neighbouring-node information. Despite restoring most of the missing third-order moment, the correction changes the macroscopic solutions considered here by less than their discretisation error: the discrepancy enters the momentum equation at Burnett order, beyond the Navier--Stokes hydrodynamic limit. The analysis also exposes a frame-dependent contribution to the viscous stress arising from Dellar's Maxwell equilibrium.

    \item \textit{The mechanisms and limits of three-dimensional robustness.}
    Three-dimensional MHD vortex simulations demonstrate second-order convergence at $\mathrm{Re}=100$ and agreement with pseudo-spectral reference solutions through the peak-current stage at $\mathrm{Re}=1000$ and, with a unit bulk relaxation rate, at $\mathrm{Re}=3000$. At $\mathrm{Re}=3000$, single-rate RR develops a compressive instability whose observed threshold depends on the grid, lattice Mach number and initial magnetic configuration. Introducing a unit bulk relaxation rate raises this threshold and suppresses the instability at $\mathrm{Ma}=0.034$ for $N\ge128$, without measurable loss of accuracy at $\mathrm{Re}=1000$; with a uniform guide field added to the initial field it does so on $N=128$ only at half that lattice Mach number. Neither regularisation nor bulk dissipation alone provides the same improvement, and the combined strategy does not prevent failure at $\mathrm{Re}=10\,000$. These results demonstrate that regularisation and bulk viscosity are distinct, complementary ingredients of robust three-dimensional MHD simulations, rather than interchangeable sources of numerical stability.

\end{itemize}
Throughout the numerical investigation, the fluid populations are treated with the collision operators under comparison, while the magnetic populations retain their BGK collision. The stability conclusions therefore concern this hybrid double-distribution formulation; they do not imply that the magnetic collision is itself optimally regularised.

The remainder of the paper is organised as follows. Section~\ref{sec:model} introduces the double-distribution model and the collision operators. Section~\ref{sec:equivalence} establishes the equivalence between RR and CHM collisions and derives the central moments of the Maxwell stress. Section~\ref{sec:ce} examines the limitations of the hydrodynamic recursion under Lorentz coupling. Section~\ref{sec:results} presents the numerical assessment of accuracy and robustness in two- and three-dimensional MHD vortex simulations. Finally, Section~\ref{sec:conclusions} summarises the findings and discusses possible extensions.


\section{The scheme and its regularisation}\label{sec:model}


This section introduces the double-distribution model and the fluid collision in four steps. We first define the tensor-product fluid lattices and the equilibrium, then the second-order non-equilibrium stress and its relaxation. We next give the recursive reconstruction of the higher-order non-equilibrium and derive its closed form. This closed form is the starting point for the equivalence and Chapman--Enskog analyses in Sections~\ref{sec:equivalence} and~\ref{sec:ce}.

The fluid populations $f_i$ and the magnetic populations $g_{j,\alpha}$ obey

\begin{equation}\label{eq:lbe}
f_i(\mathbf x+\bc_i,t+1)=f_i^{\star}(\mathbf x,t),\qquad g_{j,\alpha}(\mathbf x+\mathbf e_j,t+1)=g_{j,\alpha}^{\star}(\mathbf x,t).
\end{equation}
All quantities are expressed in lattice units, so that the time step and lattice spacing are unity. Here, $\mathbf x$ denotes a lattice node and the superscript $\star$ denotes a post-collision population. The macroscopic fields are $\rho=\sum_i f_i$, $\rho\bu=\sum_i f_i\bc_i$ and $b_\alpha=\sum_j g_{j,\alpha}$. Greek indices run over the $D$ Cartesian directions, and repeated indices are summed unless stated otherwise.

The fluid lattices D2Q9 and D3Q27 are tensor products of D1Q3. Their discrete velocities satisfy $\bc_i\in\{-1,0,1\}^D$, with components $c_{i\alpha}\in\{-1,0,1\}$. The weights are $w_i=\prod_\alpha\varpi(c_{i\alpha})$, where $\varpi(0)=2/3$ and $\varpi(\pm1)=1/6$, and the squared lattice sound speed is $c_s^2=1/3$.

The one-dimensional Hermite polynomials, defined in terms of a scalar argument \(q\), are
\[
\He_0(q)=1,\qquad
\He_1(q)=q,\qquad
\He_2(q)=q^2-c_s^2.
\]
For a multi-index $n\in\mathcal N_D=\{0,1,2\}^D$, define the multidimensional Hermite polynomials by the tensor product
\[
H_{n}(\mathbf c_i)
=\prod_{\alpha=1}^{D}\He_{n_\alpha}(c_{i\alpha}),
\qquad
|n|=\sum_\alpha n_\alpha,\qquad
n!=\prod_\alpha n_\alpha!,\qquad
\mathbf u^{n}=\prod_\alpha u_\alpha^{n_\alpha}.
\]
The $3^D$ weighted Hermite vectors $w_iH_{n}(\bc_i)$ form an orthogonal basis,
\begin{equation*}
\sum_iw_iH_{n}(\bc_i)H_{m}(\bc_i)=n!\,c_s^{2|n|}\delta_{n m},
\end{equation*}
where $\delta_{n m}$ is the Kronecker delta. Orthogonality follows from the tensor-product structure and the D1Q3 quadrature, which reproduces the Gaussian moments through fifth order; all odd moments vanish and, in particular, $\sum_c\varpi(c)c^4=c_s^2=3c_s^4$. The basis spans the independent components of the $D$-dimensional Hermite tensors $\bm{\mathcal H}^{(k)}$ (for example, $\mathcal H^{(2)}_{\alpha\beta}=c_\alpha c_\beta-c_s^2\delta_{\alpha\beta}$). Each component of $n$ is at most two, so the maximum total degree represented independently is $2D$: four on D2Q9 and six on D3Q27. Higher tensor components are aliased onto this basis; for example, $\mathcal H^{(3)}_{xxx}\equiv0$ and $\mathcal H^{(4)}_{xxxx}=-\mathcal H^{(2)}_{xx}$.

The fluid equilibrium is the sum of a product-form hydrodynamic contribution and the Maxwell-stress contribution:

\begin{equation}\label{eq:feq}
f_i^{\eq}=w_i\Big[\rho\,F_i(\bu)+\frac{M_{\alpha\beta}\mathcal H^{(2)}_{i,\alpha\beta}}{2c_s^4}\Big],\qquad
F_i(\bu)=\prod_{\alpha}F(c_{i\alpha};u_\alpha),\qquad F(c;u)=1+\frac{cu}{\cs}+\frac{(c^2-\cs)u^2}{2c_s^4}.
\end{equation}
Here, $\mathcal H^{(2)}_{i,\alpha\beta}=c_{i\alpha}c_{i\beta}-c_s^2\delta_{\alpha\beta}$ and $M_{\alpha\beta}=\tfrac12|\bb|^2\delta_{\alpha\beta}-b_\alpha b_\beta$ is the Maxwell stress, with the sign convention adopted for the momentum flux. The scalar argument $c$ in $F(c;u)$ takes the D1Q3 values $-1,0,1$; it is distinct from the vector velocity $\bc_i$. The product-form function is the complete hydrodynamic equilibrium on the represented Hermite basis~\cite{karlin2010,coreixas2017}:
\begin{equation}\label{eq:generating}
F_i(\bu)=\sum_{n\in\NN}\frac{\bu^{n}}{n!\,c_s^{2|n|}}H_{n}(\bc_i).
\end{equation}
The expansion is fourth order in the velocity on D2Q9 and includes terms up to $u_x^2u_y^2u_z^2$ on D3Q27. The Maxwell contribution supplies the magnetic part of the momentum flux~\cite{dellar2002}, while contributing neither mass nor momentum. Its trace is dimension-dependent: in three dimensions, $M_{\alpha\alpha}=|\bb|^2/2$.

The magnetic populations are defined on D2Q5 or D3Q7. The rest velocity is $\mathbf e_0=\mathbf 0$, and the $2D$ moving velocities are $\mathbf e_j=\pm\hat{\mathbf x}_\alpha$ ($j=1,\ldots,2D$). Their weights are $\psi_0$ and $\psi_{j\ne0}=(1-\psi_0)/(2D)$, with $c_{s,\mathrm m}^2=(1-\psi_0)/D$. We use $\psi_0=1/3$ in two dimensions and $\psi_0=1/4$ in three. The magnetic populations undergo a BGK collision:

\begin{equation}\label{eq:mag}
g^{\star}_{j,\alpha}=g^{\eq}_{j,\alpha}+(1-\omega_{\rm m})(g_{j,\alpha}-g^{\eq}_{j,\alpha}),\qquad
g^{\eq}_{j,\alpha}=\psi_j\Big[b_\alpha+\frac{e_{j\beta}\Lambda_{\beta\alpha}}{\csm}\Big],\qquad \Lambda_{\beta\alpha}=u_\beta b_\alpha-b_\beta u_\alpha .
\end{equation}

A Chapman--Enskog expansion recovers the incompressible MHD equations~\eqref{eq:NS_MHD_intro}--\eqref{eq:divfree_intro}, with $p=\rho c_s^2$ and reference density $\rho_0=1$, in the low-Mach-number limit~\cite{dellar2002}. The kinematic viscosity and magnetic diffusivity are $\nu=c_s^2(1/\omega-1/2)$ and $\eta=c_{s,\mathrm m}^2(1/\omega_{\rm m}-1/2)$, respectively. Here, $\omega$ is the relaxation frequency of the fluid stress and $\omega_{\rm m}$ that of the magnetic populations. The magnetic-field gradient, and hence the current, can be estimated locally~\cite{dellar2002}:

\begin{equation}\label{eq:gradb}
\Phi_{\beta\alpha}=\sum_je_{j\beta}(g_{j,\alpha}-g^{\eq}_{j,\alpha}),\qquad \partial_\beta b_\alpha=-\frac{\omega_{\rm m}}{\csm}\,\Phi_{\beta\alpha}\,[1+O(\mathrm{Ma}^2)].
\end{equation}

The second-order non-equilibrium Hermite moment is the symmetric tensor $a_{\alpha\beta}=\sum_i\mathcal H^{(2)}_{i,\alpha\beta}(f_i-f_i^\eq)$. The collision maps it to $A=\Omega_2(a)$. For a single relaxation rate, $\Omega_2(a)=(1-\omega)a$. Alternatively, the isotropic part can be assigned an independent bulk relaxation frequency $\omega_b$:

\begin{equation}\label{eq:bulk}
\Omega_2(a)=(1-\omega)\big(a-\tfrac{\operatorname{tr}a}{D}I\big)+(1-\omega_b)\tfrac{\operatorname{tr}a}{D}I ,
\end{equation}
where $I$ is the identity tensor. Thus, $\Omega_2$ relaxes the deviatoric and isotropic parts of $a$ at rates $\omega$ and $\omega_b$, respectively. RR reconstructs all higher-order non-equilibrium moments from the relaxed tensor $A$ using the hydrodynamic recursion~\cite{malaspinas2015,coreixas2017}:
\begin{equation}\label{eq:recursion}
a^{(k)}_{\alpha_1\ldots\alpha_k}=u_{\alpha_k}a^{(k-1)}_{\alpha_1\ldots\alpha_{k-1}}+\sum_{l<k}\Big(\prod_{r\ne l,k}u_{\alpha_r}\Big)A_{\alpha_l\alpha_k},\qquad a^{(2)}=A.
\end{equation}
Equivalently, the rank-$k$ tensor is a sum over all pairs of its indices. For a component indexed by $n$, there are $\binom{n_\alpha}{2}$ pairs of equal indices in direction $\alpha$, and $n_\alpha n_\beta$ mixed pairs in directions $\alpha$ and $\beta$. The corresponding coefficient in the Hermite expansion is therefore

\begin{equation}\label{eq:recop}
\RecA_{n}[A]=\sum_\alpha\binom{n_\alpha}{2}A_{\alpha\alpha}\bu^{n-2\ee_\alpha}+\sum_{\alpha<\beta}n_\alpha n_\beta A_{\alpha\beta}\bu^{n-\ee_\alpha-\ee_\beta}=\frac12A_{\alpha\beta}\frac{\partial^2\bu^{n}}{\partial u_\alpha\partial u_\beta}.
\end{equation}
Here, $\ee_\alpha$ is the unit multi-index in direction $\alpha$. The complete RR collision retains every independent Hermite component: through fourth order on D2Q9 and sixth order on D3Q27. It can be written as
\begin{equation}\label{eq:rr}
f_i^{\star}=f_i^\eq+w_i\sum_{n\in\NN}\frac{\RecA_{n}[A]}{n!\,c_s^{2|n|}}H_{n}(\bc_i).
\end{equation}
The differential representation in~\eqref{eq:recop} gives a compact form of this reconstruction. Applying $\tfrac12A_{\alpha\beta}\partial^2/(\partial u_\alpha\partial u_\beta)$ term by term to the generating expansion~\eqref{eq:generating} yields
\begin{equation}\label{eq:closed}
f_i^{\star}=f_i^\eq+R_i(A),\qquad R_i(A)=\frac{w_i}{2}A_{\alpha\beta}\frac{\partial^2F_i}{\partial u_\alpha\partial u_\beta}.
\end{equation}
For a repeated direction, $\partial^2F_i/\partial u_\alpha^2=F''(c_{i\alpha};u_\alpha)\prod_{\gamma\ne\alpha}F(c_{i\gamma};u_\gamma)$; for distinct directions, $\partial^2F_i/\partial u_\alpha\partial u_\beta=F'(c_{i\alpha};u_\alpha)F'(c_{i\beta};u_\beta)\prod_{\gamma\ne\alpha,\beta}F(c_{i\gamma};u_\gamma)$, where $F'(c;u)=c/c_s^2+(c^2-c_s^2)u/c_s^4$ and $F''(c;u)=(c^2-c_s^2)/c_s^4$.

This discrete identity is the counterpart of a familiar Maxwellian identity. For a Maxwellian $f^{\rm M}$ at temperature $\theta$,
\begin{equation}
\tfrac12A_{\alpha\beta}\,\frac{\partial^2(f^{\rm M}/\rho)}{\partial u_\alpha\partial u_\beta}
=(f^{\rm M}/\rho)\,\frac{A_{\alpha\beta}(\xi_\alpha\xi_\beta-\theta\delta_{\alpha\beta})}{2\theta^2},
\qquad \boldsymbol\xi=\bc-\bu,
\end{equation}
which, for traceless $A$, gives the stress contribution in Grad's thirteen-moment distribution~\cite{grad1949,shan2006}. On the tensor-product lattices considered here, the product-form equilibrium reproduces the Maxwellian identity exactly on the independent Hermite components. The closed form replaces the 17 higher-order recursion coefficients used for D3Q27 in~\cite{malaspinas2015} with the one-dimensional factors $F$, $F'$ and $F''$, while reproducing the original reconstruction to round-off (Appendix~\ref{app:checks}). Restricting the reconstruction to independent components is essential: including aliased components in a full tensor contraction feeds them back into lower-order moments. For example, on D2Q9, $\mathcal H^{(4)}_{xxxx}=-\mathcal H^{(2)}_{xx}$ makes the reconstructed $a_{xx}$ contribution acquire a factor $1-4.5u_x^2$ (Section~\ref{sec:res2d}).


\section{Recursive regularisation is a central-moment collision}\label{sec:equivalence}


For a population vector $f$, define the central Hermite moments by
\begin{equation}\label{eq:central}
K_{n}(f;\bu)=\sum_if_i\prod_\alpha\He_{n_\alpha}(c_{i\alpha}-u_\alpha),\qquad n\in\NN.
\end{equation}
These moments are evaluated in the local comoving frame. The transformation $f\mapsto K(f;\bu)$ is a Kronecker product of one-dimensional $3\times3$ transformations. Each factor is a Vandermonde transformation evaluated at the three shifted nodes $c-u$ (followed by a unit lower-triangular change of polynomial basis); its determinant is non-zero and independent of $u$. The transformation is therefore invertible for every value of the comoving velocity. A central-Hermite-moment (CHM) collision conserves the zeroth- and first-order moments, applies $\Omega_2$ to the second-order non-equilibrium moments, and sets all higher-order central Hermite moments to their equilibrium values:

\begin{equation}\label{eq:chm}
K^\star_{n}=K_{n}(f^\eq;\bu)\ \ (|n|\ne2),\qquad K^\star_{\nab}=K_{\nab}(f^\eq;\bu)+[\Omega_2(S)]_{\alpha\beta},\qquad S_{\alpha\beta}=K_{\nab}(f-f^\eq;\bu),
\end{equation}

Here, $\nab=\ee_\alpha+\ee_\beta$, and the post-collision populations are obtained by inverting~\eqref{eq:central}. Since $f$ and $f^\eq$ have the same density and momentum, the second-order central moment of $f-f^\eq$ equals its second-order raw Hermite moment; hence $S=a$.

The following equivalence is the central result of this section. On D2Q9 and D3Q27, the CHM collision~\eqref{eq:chm} and the complete RR reconstruction~\eqref{eq:closed} produce identical post-collision populations, provided that they use the same comoving velocity, equilibrium and second-order relaxation map $\Omega_2$. The result holds for any pre-collision state and does not require $\Omega_2$ to be linear. The product-form function $F_i$ enters only through the reconstruction $R_i$; it does not constrain the equilibrium used by the collision, beyond the requirement that this equilibrium have the prescribed density and momentum.

The proof follows from the central moments of $R_i(A)$. In one dimension, define $\mu_p[h]=\sum_c\varpi(c)h(c;u)\He_p(c-u)$. Direct evaluation using the D1Q3 quadrature identity $\sum_c\varpi(c)c^4=3c_s^4$ gives $\mu_p[F]=\delta_{p0}$. Differentiating with respect to $u$ and using $\He_p'=p\He_{p-1}$ gives $\mu_p[\partial_u h]=\partial_u\mu_p[h]+p\mu_{p-1}[h]$, and therefore $\mu_p[F']=\delta_{p1}$ and $\mu_p[F'']=2\delta_{p2}$. Since $R_i(A)$ is assembled from products of these one-dimensional factors, its central Hermite moments vanish at every order except the second, where they equal $A_{\alpha\beta}$. Consequently, the CHM prescription gives $K^\star=K(f^\eq;\bu)+K(R(\Omega_2(a)))$. Inverting the central-moment transformation yields $f^\star=f^\eq+R(\Omega_2(a))$, which is exactly~\eqref{eq:closed}. In other words, RR constructs the unique population vector whose non-equilibrium central Hermite moments vanish at all orders except second order, where they are given by the relaxed tensor $A$.

This result extends the equivalence established in~\cite{coreixas2019} for a single relaxation rate and the isothermal equilibrium. In that work, the equivalence is lost when multiple relaxation times are applied to a reconstructed non-equilibrium through a relaxation matrix. Here, by contrast, the recursion acts on the already relaxed tensor $A=\Omega_2(a)$, so the equivalence holds for any common second-order map $\Omega_2$. This includes a separate bulk rate~\eqref{eq:bulk}, general relaxation matrices and state-dependent rates, such as those used in eddy-viscosity models. The same argument also extends to a body force $\mathbf s$ when the velocity is defined using Guo's half-force shift, $\bu=(\sum_i f_i\bc_i+\mathbf s/2)/\rho$~\cite{guo2002}. The force contributes to the first-order central moment, with post-collision value $K^\star_{\ee_\alpha}=s_\alpha/2$, as in established central-moment forcing schemes~\cite{premnath2009,fei2017}:
\begin{equation}\label{eq:force}
f_i^\star=f_i^\eq+R_i\big(\Omega_2(a+\tfrac12(\bu\mathbf s+\mathbf s\bu))\big)+\tfrac12G_i(\mathbf s),\qquad G_i(\mathbf s)=w_is_\gamma\frac{\partial F_i}{\partial u_\gamma}.
\end{equation}
Here, $G_i$ is the forcing term expanded over the independent Hermite components. Its only non-zero central Hermite moment is the first-order contribution $s_\gamma$; the prefactor $1/2$ therefore gives the required post-collision first-order moment. For comparison, projective regularisation~\cite{latt2006} is defined by
\begin{equation}\label{eq:pr}
f_i^\star=f_i^\eq+\frac{w_i}{2c_s^4}A_{\alpha\beta}\mathcal H^{(2)}_{i,\alpha\beta}.
\end{equation}
Equation~\eqref{eq:pr} is the corresponding raw-Hermite-moment collision. Its difference from RR is
\begin{equation*}
w_i\sum_{|n|\ge3}\frac{\RecA_{n}[A]}{n!\,c_s^{2|n|}}H_{n}(\bc_i),
\end{equation*}
which is generally non-zero at finite velocity. The distinction is therefore the frame in which the non-equilibrium is filtered: projective regularisation acts in the laboratory frame, whereas RR performs the filtering in the local comoving frame (Fig.~\ref{fig:spectra}).

\begin{figure}[!htbp]

\centering

\begin{tikzpicture}
\begin{semilogyaxis}[width=0.58\linewidth, height=5.6cm, ybar=0.6pt, bar width=5.2pt, log origin=infty,
  xtick={2,3,4,5,6}, enlarge x limits=0.14, ymin=1e-21, ymax=3e-2,
  ytick={1e-20,1e-16,1e-12,1e-8,1e-4},
  xlabel={order $k$ of the central Hermite moments}, ylabel={$\big(\sum_{|n|=k}K_n^2\big)^{1/2}$},
  legend style={at={(1.02,1)}, anchor=north west, draw=none, font=\footnotesize, cells={anchor=west}}, legend cell align=left]
\addplot[fill=cBGK!70, draw=cBGK] table[x=order, y=BGK] {data/spectra.dat};
\addplot[fill=cPR!80, draw=cPR!80!black] table[x=order, y=PR] {data/spectra.dat};
\addplot[fill=cRR!85, draw=cRR!80!black] table[x=order, y=RR] {data/spectra.dat};
\addplot[fill=cRRL!80, draw=cRRL!70!black] table[x=order, y=RRL] {data/spectra.dat};
\legend{BGK (first-order CE), projective reg., recursive reg., RR + Lorentz corr.}
\end{semilogyaxis}
\end{tikzpicture}

\caption{Central Hermite moments of the post-collision non-equilibrium, grouped by order $k$, for one D3Q27 state: $\rho=1$, $\bu=(0.06,-0.04,0.03)$, $\bb=(0.05,0.03,-0.04)$, prescribed gradients of $\bu$ and $\bb$, and $\omega=1.9$. The pre-collision non-equilibrium is the first-order Chapman--Enskog solution~\eqref{eq:f1} of the MHD equilibrium~\eqref{eq:feq}, which the BGK bars show. Recursive regularisation leaves central moments at second order only (round-off beyond); projective regularisation, which filters in the frame at rest, leaves spurious higher-order ones. The Lorentz correction~\eqref{eq:df} (level 1) restores the Chapman--Enskog values: to $0.02\%$ at orders $3$ and $4$, and to $2\%$ at orders $5$ and $6$, where the $O(\mathrm{Ma}^2)$ terms are missing.}

\label{fig:spectra}

\end{figure}
The same calculation gives the central moments of the full MHD equilibrium~\eqref{eq:feq}. These are the equilibrium values that must be prescribed by a CHM implementation:
\begin{equation}\label{eq:maxwellK}
K_{n}(f^\eq;\bu)=\rho\,\delta_{n,\mathbf 0}+(-1)^{|n|}\,\RecA_{n}[M].
\end{equation}

The hydrodynamic contribution is $\rho\prod_\alpha\mu_{n_\alpha}[F]=\rho\delta_{n,\mathbf 0}$. For the Maxwell contribution, applying the translation identity $\He_p(c-u)=\sum_{q\le p}\binom pq(-u)^{p-q}\He_q(c)$ and Hermite orthogonality gives the stated result. Thus, up to the sign $(-1)^{|n|}$, the central Hermite moments of the Maxwell contribution are precisely the coefficients generated by the RR reconstruction with $A=M$. They vanish at orders zero and one, equal $M_{\alpha\beta}$ at second order, and are generally non-zero at higher represented orders. For reference, the corresponding equilibrium monomial central moments on D2Q9 are
\begin{gather}
k_{21}=\tfrac12u_y(b_x^2-b_y^2)+2u_xb_xb_y,\qquad k_{12}=\tfrac12u_x(b_y^2-b_x^2)+2u_yb_xb_y,\nonumber\\
k_{22}=\rho c_s^4+\tfrac12(b_x^2-b_y^2)(u_x^2-u_y^2)-4u_xu_yb_xb_y .\label{eq:k2d}
\end{gather}
These are the equilibrium moments of the D2Q9 central-moment scheme in~\cite{derosis2021}. They lead to three practical conclusions. First, a CHM implementation using the full equilibrium~\eqref{eq:feq} must include the Maxwell contribution in its equilibrium moments through the highest represented order, as specified by~\eqref{eq:maxwellK}. If instead the Maxwell stress is retained only at second order and all higher-order Maxwell central Hermite moments are set to their hydrodynamic equilibrium values, the resulting collision is RR with the comoving equilibrium
\begin{equation}\label{eq:comoving}
f_i^{\eq\prime}=w_i\rho F_i(\bu)+R_i(M)=w_i\Big[\rho F_i(\bu)+\tfrac12M_{\alpha\beta}\frac{\partial^2F_i}{\partial u_\alpha\partial u_\beta}\Big].
\end{equation}
With this choice, the recursion propagates the Maxwell stress through the Hermite hierarchy in the same way that it propagates the hydrodynamic pressure. For example, its third-order coefficient is $u_\alpha M_{\beta\gamma}+u_\beta M_{\alpha\gamma}+u_\gamma M_{\alpha\beta}$. Both equilibrium choices are consistent with the equivalence above. The comoving form is simpler to implement in central-moment space; the two formulations differ appreciably only when a strong magnetic field drives the flow to large lattice velocities (Section~\ref{sec:res3d}). Second, in raw Hermite moments the Maxwell contribution has only a second-order component, so a raw-Hermite or projective scheme requires no higher-order Maxwell moments. Third, in raw monomial moments the Maxwell contribution also appears at higher orders: it contributes $\cs(M_{\alpha\alpha}+M_{\beta\beta})$ at $2\ee_\alpha+2\ee_\beta$, $\cs M_{\alpha\beta}$ at $\ee_\alpha+\ee_\beta+2\ee_\gamma$, and $c_s^4\operatorname{tr}M$ at $(2,2,2)$. These terms vanish on D2Q9, where $\operatorname{tr}M=0$, but not generally on D3Q27.

Many central-moment schemes are formulated using monomial central moments $k_n$~\cite{geier2006,derosis2018,derosis2021}. Hermite and monomial central moments are related by the unit lower-triangular transformation $K=\Pi k$, where $\Pi=\Pi_1^{\otimes D}$. This follows from $\He_2(C)=C^2-c_s^2$, with $C=c-u$; the diagonal entries of $\Pi_1$ are unity and its $(2,0)$ entry is $-c_s^2$. Thus, setting higher-order monomial moments to equilibrium is not equivalent to setting higher-order Hermite moments to equilibrium. A monomial collision that relaxes the second-order moments according to $\Omega_2$ produces $k^\eq+\iota(A)$, where $k^\eq$ denotes the monomial central moments of $f^\eq$ and $\iota(A)$ inserts $A$ into the second-order slots. In Hermite coordinates this becomes $K^\eq+\Pi\iota(A)$, rather than $K^\eq+\iota(A)$. The discrepancy $(\Pi-I)\iota(A)$ is
\begin{equation}\label{eq:monodiff}
\Delta K_{n}=\begin{cases}
-\cs(A_{xx}+A_{yy}), & n=(2,2)\ \text{on D2Q9},\\
-\cs(A_{\alpha\alpha}+A_{\beta\beta}), & n=2\ee_\alpha+2\ee_\beta\ \text{on D3Q27},\\
-\cs A_{\alpha\beta}, & n=\ee_\alpha+\ee_\beta+2\ee_\gamma\ \text{on D3Q27},\ \alpha,\beta,\gamma\ \text{distinct},\\
c_s^4\operatorname{tr}A, & n=(2,2,2),
\end{cases}
\end{equation}
and all other components vanish. On D2Q9, the monomial central-moment collision and RR therefore coincide when $\operatorname{tr}A=0$, in particular when the trace is relaxed to equilibrium ($\omega_b=1$). This is consistent with the relationship between cascaded and multiple-relaxation-time RR models reported in~\cite{coreixas2019}. Applying the same argument to raw moments (setting the shift $\bu$ to zero in the transformation) shows that, on D2Q9, a raw-monomial collision with higher-order moments at equilibrium is projective regularisation when $\operatorname{tr}A=0$, namely $\mathrm{PR}_{\omega,1}$.

On D3Q27, however, deviatoric components of $A$ also enter fourth-order moments, and no choice of bulk relaxation rate removes the discrepancy. This is not an intrinsic limitation of the monomial basis: Hermite and monomial central moments span the same space. A monomial collision would reproduce RR exactly if its higher-order moments were set to $k^\eq+(\Pi^{-1}-I)\iota(A)$ rather than simply to $k^\eq$. The difference lies in which higher-order quantities are constrained to equilibrium. Through the lower-triangular transformation $\Pi$, setting the fourth-order monomial moments to equilibrium leaves a contribution proportional to the relaxed stress $A$ in the fourth-order Hermite moments, including its deviatoric part; the CHM collision removes that contribution. This explains the loss of accuracy measured in Section~\ref{sec:res3d}. A monomial central-moment implementation using the comoving equilibrium~\eqref{eq:comoving} must likewise retain the higher-order Maxwell contributions identified above. The collision operators considered in this work are summarised in Table~\ref{tab:operators}.

\begin{table}[!htbp]

\centering\small

\caption{Collision operators considered in this work and their equivalences. All use the equilibrium~\eqref{eq:feq} unless stated otherwise. $\mathrm{RR}_{\omega,\omega_b}$ and $\mathrm{PR}_{\omega,\omega_b}$ denote recursive and projective regularisation with the map~\eqref{eq:bulk}, and $\mathrm{BGK}_{\omega,\omega_b}$ BGK with the trace of the non-equilibrium relaxed at $\omega_b$.}

\label{tab:operators}

\begin{tabular}{@{}p{3.7cm}p{4.3cm}p{2.5cm}p{3.6cm}@{}}

\toprule

operator & moments & orders $\ge3$ & equivalent to\\

\midrule

BGK & -- & relaxed at $\omega$ & --\\

$\mathrm{BGK}_{\omega,\omega_b}$ (D3Q27) & -- & relaxed at $\omega$ & --\\

RR (D2Q9, D3Q27) & -- & recursion~\eqref{eq:rr} & CHM with the same $\Omega_2$\\

PR (D2Q9, D3Q27) & -- & discarded & raw Hermite, higher orders at equilibrium\\

CHM & central Hermite & equilibrium & RR with the same $\Omega_2$\\

CMs (D2Q9) & central monomial, trace to equilibrium & equilibrium & $\mathrm{RR}_{\omega,1}$\\

RMs (D2Q9) & raw monomial, trace to equilibrium & equilibrium & $\mathrm{PR}_{\omega,1}$\\

CM (D3Q27) & central monomial, trace to equilibrium & equilibrium & none (differs at fourth order)\\

\bottomrule

\end{tabular}

\end{table}


\section{What the hydrodynamic recursion misses}\label{sec:ce}


RR reconstructs the higher-order non-equilibrium as if the flow were purely hydrodynamic. To quantify the resulting discrepancy, we compare the reconstruction with the first-order Chapman--Enskog (CE) non-equilibrium of the MHD equilibrium,
$f_i^{(1)}=-\tau(\partial_t^{(1)}+c_{i\gamma}\partial_\gamma)f_i^\eq$, where $\tau=1/\omega$ and $\partial_t^{(1)}$ is the first-order time derivative in the multiple-scale expansion. The derivation in this section uses a single fluid relaxation rate, so that the second-order non-equilibrium is relaxed with $\omega_b=\omega$. We use the ideal continuity, momentum and induction equations,
$\partial_t\rho+\nabla\cdot(\rho\bu)=0$,
$\rho(\partial_t+\bu\cdot\nabla)u_\alpha=-\partial_\alpha(\rho c_s^2)-\partial_\beta M_{\alpha\beta}$ and
$\partial_tb_\alpha+\partial_\beta\Lambda_{\beta\alpha}=0$,
together with the D1Q3 Hermite recurrence. The resulting Hermite coefficients, for $k\ge2$ (Appendix~\ref{app:ce}), are
\begin{equation}\label{eq:a1}
a^{(k)}_{1,\alpha_1\ldots\alpha_k}=\sum_{l<m}a^{\rm h}_{\alpha_l\alpha_m}\prod_{r\ne l,m}u_{\alpha_r}+\tau\sum_{l=1}^k\partial_\beta M_{\alpha_l\beta}\prod_{r\ne l}u_{\alpha_r}-\tau\,\delta_{k2}\,\partial_tM_{\alpha_1\alpha_2}-\tau\cs\,\delta_{k3}\,T_{\alpha_1\alpha_2\alpha_3}.
\end{equation}
Here, the subscript $1$ in $a_1^{(k)}$ denotes first order in the CE expansion, not Hermite order. The hydrodynamic contribution is $a^{\rm h}_{\alpha\beta}=-\tau\rho c_s^2(\partial_\alpha u_\beta+\partial_\beta u_\alpha)$, and $T_{\alpha\beta\gamma}=\partial_\alpha M_{\beta\gamma}+\partial_\beta M_{\alpha\gamma}+\partial_\gamma M_{\alpha\beta}$. The four terms in~\eqref{eq:a1} represent, in order, the hydrodynamic non-equilibrium, the Lorentz-force contribution, the time derivative of the Maxwell stress and the streaming contribution from the Maxwell part of the equilibrium. The same result can be expressed directly in population space:
\begin{equation}\label{eq:f1}
f_i^{(1)}=R_i(a^{\rm h})+\tau w_i\,\partial_\beta M_{\gamma\beta}\Big[\frac{\partial F_i}{\partial u_\gamma}-\frac{c_{i\gamma}}{\cs}\Big]-\frac{\tau w_i}{2c_s^4}\big[\mathcal H^{(2)}_{i,\alpha\beta}\partial_tM_{\alpha\beta}+\mathcal H^{(3)}_{i,\alpha\beta\gamma}\partial_\gamma M_{\alpha\beta}\big].
\end{equation}
Equation~\eqref{eq:f1} is exact for the independent Hermite components, except for components with a repeated index ($n_\gamma=2$), which also contain the cubic hydrodynamic-stress defect $\tau\sum_{\gamma:\,n_\gamma=2}\partial_\gamma(\rho\bu^{n} u_\gamma)$. This exception is a consequence of D1Q3 aliasing: the corresponding third-order one-dimensional Hermite polynomial vanishes on the lattice.

At second order, $a^{(2)}_1=a^{\rm h}+\mathcal L$ with
\begin{equation}\label{eq:L}
\mathcal L_{\alpha\beta}=\tau\big[u_\alpha\partial_\gamma M_{\beta\gamma}+u_\beta\partial_\gamma M_{\alpha\gamma}-\partial_tM_{\alpha\beta}\big].
\end{equation}
The momentum equation therefore contains the viscous stress $(1-\omega/2)(a^{\rm h}+\mathcal L)$. The additional term originates in the equilibrium: Dellar's Maxwell contribution contains no third-order Hermite coefficients. For characteristic magnetic and velocity amplitudes of comparable magnitude and comparable spatial scales, $\mathcal L$ is of relative order $O(\mathrm{Ma}^2)$ compared with the hydrodynamic viscous stress. This is comparable to the cubic defect already known in hydrodynamic lattice Boltzmann schemes~\cite{dellar2014}. However, $\mathcal L$ depends explicitly on the velocity. Under a change of reference frame with velocity $\mathbf U$, the term $u_\alpha\partial_\gamma M_{\beta\gamma}$ acquires a contribution of characteristic magnitude $|\mathbf U||\bb||\nabla\bb|$; the viscous stress generated by Dellar's equilibrium is therefore frame dependent. With the comoving equilibrium~\eqref{eq:comoving}, the third-order Maxwell coefficient contributes $-\tau\partial_\gamma(u_\alpha M_{\beta\gamma}+u_\beta M_{\alpha\gamma}+u_\gamma M_{\alpha\beta})$. Combining this with~\eqref{eq:L} gives
\begin{equation*}
 a^{(2)}_{1,\alpha\beta}-a^{\rm h}_{\alpha\beta}
 =-\tau\left[M_{\beta\gamma}\partial_\gamma u_\alpha
 +M_{\alpha\gamma}\partial_\gamma u_\beta
 +M_{\alpha\beta}\nabla\cdot\bu
 +(\partial_t+\bu\cdot\nabla)M_{\alpha\beta}\right].
\end{equation*}
Using the ideal induction equation and $\nabla\cdot\bb=0$ to evaluate the material derivative of $M_{\alpha\beta}$ then yields
\begin{multline}\label{eq:R}
a^{(2)}_1=a^{\rm h}-\tau\Big[\delta_{\alpha\beta}\big(\bb\cdot\nabla\bu\cdot\bb-\tfrac12|\bb|^2\nabla\cdot\bu\big)+b_\alpha b_\beta\nabla\cdot\bu\\
+\tfrac12|\bb|^2(\partial_\alpha u_\beta+\partial_\beta u_\alpha)-2\big(b_\beta(\bb\cdot\nabla)u_\alpha+b_\alpha(\bb\cdot\nabla)u_\beta\big)\Big].
\end{multline}
The Maxwell contribution remaining in~\eqref{eq:R} depends on $\bb\otimes\bb$ and $\nabla\bu$, but not explicitly on the frame velocity. Its characteristic magnitude relative to the hydrodynamic viscous stress is $O(|\bb|^2/(\rho c_s^2))$. Removing it would require the full velocity gradient, which is not available from the local information retained by the scheme. Moreover, the discrete lattice cannot represent all of the required higher-order tensor components independently; for example, the $xxx$ component contains an aliased contribution proportional to $3u_xM_{xx}$. The normal stresses can therefore retain a frame-dependent discrete defect, analogous to that found in hydrodynamic lattice Boltzmann schemes~\cite{dellar2014}. The frame-independent continuum contribution and this discrete aliasing error are distinct effects.

A separate discrepancy occurs at third order. Since the measured second-order CE coefficient $a_1^{(2)}$ already contains $\mathcal L$, the RR reconstruction agrees with the CE result at second order. At higher orders, however, the recursion~\eqref{eq:recursion} propagates $\mathcal L$ by multiplying it by velocity components. The Lorentz-force contribution is consequently counted $k-1$ times in the reconstructed rank-$k$ tensor, rather than once, and the time derivative $\partial_tM$ is propagated to all higher orders. Defining the difference as $\Delta a^{(k)}=a_{\rm CE}^{(k)}-a_{\rm RR}^{(k)}$, we obtain, for $k\ge3$,
\begin{equation}\label{eq:delta}
\Delta a^{(k)}_{\alpha_1\ldots\alpha_k}=-\tau\cs\,\delta_{k3}\,T_{\alpha_1\alpha_2\alpha_3}-\tau(k-2)\sum_l\partial_\beta M_{\alpha_l\beta}\prod_{r\ne l}u_{\alpha_r}+\tau\sum_{l<m}\partial_tM_{\alpha_l\alpha_m}\prod_{r\ne l,m}u_{\alpha_r}.
\end{equation}
The first term in~\eqref{eq:delta} is the leading contribution omitted by the hydrodynamic recursion. Relative to a characteristic hydrodynamic third-order coefficient, its magnitude scales as $|\bb||\nabla\bb|/(\rho|\bu||\nabla\bu|)$. It can therefore be order one when the magnetic field is not weak, and is potentially important in current sheets or in the presence of a guide field. This is a characteristic-magnitude estimate and is not meaningful where the hydrodynamic reference contribution vanishes. The remaining two terms are smaller by $O(\mathrm{Ma}^2)$. The leading discrepancy also occurs with the comoving equilibrium: its Maxwell contribution enters through the same streaming term, $c_s^2\sum_\gamma n_\gamma\partial_\gamma M_{n-\ee_\gamma}$ in~\eqref{eq:a1n}, where $M_m$ denotes the Hermite coefficient of the Maxwell part of the equilibrium with multi-index $m$. We therefore introduce a local correction that restores this leading third-order contribution:
\begin{equation}\label{eq:df}
\Delta f_i=\frac{w_i}{6c_s^6}\Delta\bm a^{(3)}\!:\!\bm{\mathcal H}^{(3)}_i=-\frac{\tau w_i}{2c_s^4}\mathcal H^{(3)}_{i,\alpha\beta\gamma}\partial_\gamma M_{\alpha\beta},\qquad
\partial_\gamma M_{\alpha\beta}=\delta_{\alpha\beta}b_\delta\partial_\gamma b_\delta-b_\beta\partial_\gamma b_\alpha-b_\alpha\partial_\gamma b_\beta .
\end{equation}
The magnetic-field gradient in~\eqref{eq:df} is evaluated using~\eqref{eq:gradb}, so the correction is local and requires no neighbouring-node data. Because it contains only a third-order Hermite contribution, it leaves the zeroth-, first- and second-order raw Hermite moments unchanged. In central-moment space, the third-order post-collision contribution is $-(1-\omega)\tau c_s^2T_{n}$, which follows when the correction is applied before the fluid collision. The insertion stage should therefore be stated explicitly and checked against the implementation. At higher orders, the velocity shift also generates the contributions $(1-\omega)(-1)^{|n|-3}\sum_{|\boldsymbol m|=3}\prod_\alpha\binom{n_\alpha}{m_\alpha}u_\alpha^{n_\alpha-m_\alpha}\Delta a_{\boldsymbol m}$, where $\Delta a_{\boldsymbol m}$ are the components of $\Delta\bm a^{(3)}$.

The discrepancy does not alter the macroscopic equations at Navier--Stokes order. Three levels of the argument should be distinguished. First, the error in the third-order moment itself can be leading order: the hydrodynamic recursion may miss a contribution comparable to the moment it reconstructs. Second, the third-order moment enters the evolution of the second-order stress through streaming, via its divergence. Since the post-collision third-order non-equilibrium is proportional to $(1-\omega)a_1^{(3)}$, its effect on the stress appears at second order in the Knudsen expansion, beyond the Navier--Stokes stress. Third, the momentum equation contains the divergence of that stress, so the resulting defect enters at third order, i.e. at Burnett order, which is also the order of the scheme's truncation error. A large relative error in a third-order moment is therefore compatible with a small change in the macroscopic solution; both effects are observed in the simulations (Appendix~\ref{app:lorentz}). Using BGK populations, which retain the underlying non-equilibrium distribution, we find that the hydrodynamic recursion reconstructs the third-order non-equilibrium with a relative error of $78$--$96\%$ in the tested MHD states, compared with about $4\%$ without a magnetic field. The local correction reduces this error to $2$--$4\%$ at $t\approx0.2$. However, for the standard field at $\mathrm{Re}=100$ on $N=64$, the corrected and uncorrected runs differ by at most $0.035\%$ in $E_{\rm k}$, $0.017\%$ in $E_{\rm m}$ and $0.11\%$ in $J_{\max}$; these differences are between five and twenty-five times smaller than the respective errors relative to the spectral reference ($0.20\%$, $0.32\%$ and $2.7\%$). The difference also decreases faster than the reference errors under grid refinement (Appendix~\ref{app:lorentz}). The correction does not improve stability. For the macroscopic MHD predictions considered here, the uncorrected hydrodynamic recursion is therefore sufficient.

The magnetic populations require a separate treatment. D2Q5 and D3Q7 are not tensor products of D1Q3, so the RR--CHM equivalence established in Section~\ref{sec:equivalence} does not apply to them. For each magnetic-field component, projective regularisation of~\eqref{eq:mag} retains the first-order non-equilibrium moment $\Phi$ and removes the raw second-order moments. A comoving variant additionally retains $(1-\omega_{\rm m})2u_\beta\Phi_{\beta\alpha}$. Neither reconstruction reproduces the second-order CE moment retained by BGK:

\begin{equation}\label{eq:Nce}
\sum_je_{j\beta}^2(g_{j,\alpha}-g^\eq_{j,\alpha})=-\tau_{\rm m}\big[\partial_\beta\Lambda_{\beta\alpha}-\csm\partial_\gamma\Lambda_{\gamma\alpha}\big]\qquad(\text{no sum over }\beta),
\end{equation}
Here, $\tau_{\rm m}=1/\omega_{\rm m}$, and there is no sum over $\beta$ in the left-hand side or in the first term on the right-hand side. To identify the information required by this moment, expand the flux derivatives:
\begin{align*}
\partial_\beta\Lambda_{\beta\alpha}
&=(\partial_\beta u_\beta)b_\alpha
 +u_\beta\partial_\beta b_\alpha
 -(\partial_\beta b_\beta)u_\alpha
 -b_\beta\partial_\beta u_\alpha,\qquad(\text{no sum over }\beta),\\
\partial_\gamma\Lambda_{\gamma\alpha}
&=(\nabla\cdot\bu)b_\alpha
 +u_\gamma\partial_\gamma b_\alpha
 -(\nabla\cdot\bb)u_\alpha
 -b_\gamma\partial_\gamma u_\alpha,
\end{align*}
where $\gamma$ is summed. Thus, even after imposing the solenoidal constraints, the fixed-direction term contains components of the velocity gradient, including $\partial_\beta u_\beta$ and $\partial_\beta u_\alpha$, as well as magnetic-field gradients. The local magnetic non-equilibrium moment $\Phi$ provides an estimate of $\partial_\beta b_\alpha$ through~\eqref{eq:gradb}, but does not provide these velocity-gradient components. The limitation is therefore not, in general, attributable solely to the antisymmetric part of $\nabla\bu$: the full second-order CE moment requires directional velocity-gradient information that is absent from the local magnetic non-equilibrium retained by these regularisations. Its contribution enters the induction equation only at third order in the Knudsen number. We therefore retain the BGK collision for the magnetic populations throughout this study.

\section{Numerical evidence}\label{sec:results}

All schemes use the same streaming and the same BGK collision for the magnetic populations. In two dimensions they differ only in the fluid collision: BGK; raw-moment (RMs) and central-moment (CMs) schemes with monomial moments, the trace relaxed to equilibrium and the higher orders at equilibrium~\cite{derosis2021}; RR with a single rate; and RR and projective regularisation with $\omega_b=1$. In three dimensions RR uses the closed form~\eqref{eq:closed} with the complete equilibrium and, optionally, the correction~\eqref{eq:df}; projective regularisation and BGK with the trace of the non-equilibrium relaxed at its own rate serve as controls. The runs of Figs.~\ref{fig:re1000}--\ref{fig:highre} and~\ref{fig:iso3dgpu}, and the entries of Table~\ref{tab:div} on $N\ge256$, with those of BGK and RR on $N=128$ at $\mathrm{Re}=1000$, use the CHM collision, started from the same equilibrium. By Section~\ref{sec:equivalence} it is the same scheme as RR, and it agrees with the RR implementation to round-off over a whole run (\ref{app:checks}).

The two-dimensional Orszag--Tang vortex~\cite{orszag1979} starts from $\bu=u_0(-\sin y,\sin x)$ and $\bb=b_0(-\sin y,\sin2x)$ with $u_0=b_0=2$ in $[0,2\pi)^2$, on $N^2$ nodes, with $\mathrm{Re}=u_0L/\nu$, $L=2\pi$, magnetic Prandtl number $\mathrm{Pr_m}=\nu/\eta=1$ and the lattice velocity amplitude $\tilde u_0=1/(20\pi)$, i.e.\ $\mathrm{Ma}=\tilde u_0/c_s\approx0.028$, $\Delta x=2\pi/N$ and $\Delta t=1/(20N)$. The three-dimensional vortex mimics the setup proposed by Mininni \textit{et al.}~\cite{mininni2006}, where the flow starts from
\begin{equation}
\bu=2\,(-\sin y,\sin x,0),\qquad \bb=0.8\,(-2\sin2y+\sin z,\;2\sin x+\sin z,\;\sin x+\sin y)
\end{equation}
in $[0,2\pi)^3$ on $N^3$ nodes, with $\nu=\eta=2\pi/\mathrm{Re}$, lattice velocity $v_0=6.94\times10^{-3}$ for unit physical speed ($\mathrm{Ma}=2\sqrt2\,v_0/c_s=0.034$ on the peak initial speed $2\sqrt2$). Both flows start from uniform density with all populations at equilibrium. The references are pseudo-spectral, with $N_s$ Fourier modes per direction (\ref{app:spectral}). With the current density $\mathbf j=\nabla\times\bb$, we report the peak current $j_{\max}=\max|j_z|$ in two dimensions, obtained locally from~\eqref{eq:gradb}, and the kinetic and magnetic energies $E_{\rm k}$ and $E_{\rm m}$ and the peak current $J_{\max}=\max|\nabla\times\bb|$ in three, $J_{\max}$ by central differences, compared with the same stencil applied to the spectral solution on the lattice nodes; the lower-case and the capital symbol distinguish the two definitions.

\subsection{Two dimensions}\label{sec:res2d}

\begin{figure}[!htbp]
\centering
\begin{tikzpicture}
\begin{groupplot}[group style={group size=2 by 2, horizontal sep=2.4cm, vertical sep=2.2cm},
  scale only axis, width=5.0cm, height=5.0cm, enlargelimits=false, axis on top,
  xmin=0, xmax=6.28319, ymin=0, ymax=6.28319,
  xtick={0,3.14159,6.28319}, xticklabels={$0$,$\pi$,$2\pi$}, ytick={0,3.14159,6.28319}, yticklabels={$0$,$\pi$,$2\pi$},
  xlabel={$x$}, ylabel={$y$}, ylabel style={yshift=-0.2cm},
  colorbar, colormap name=rdbu, colorbar style={width=0.18cm, font=\scriptsize},
  title style={font=\small}]
\nextgroupplot[title={(a) $\mathrm{Re}=200\pi$, $t=0.5$}, point meta min=-18.5, point meta max=18.5]
\addplot graphics[xmin=0, xmax=6.28319, ymin=0, ymax=6.28319] {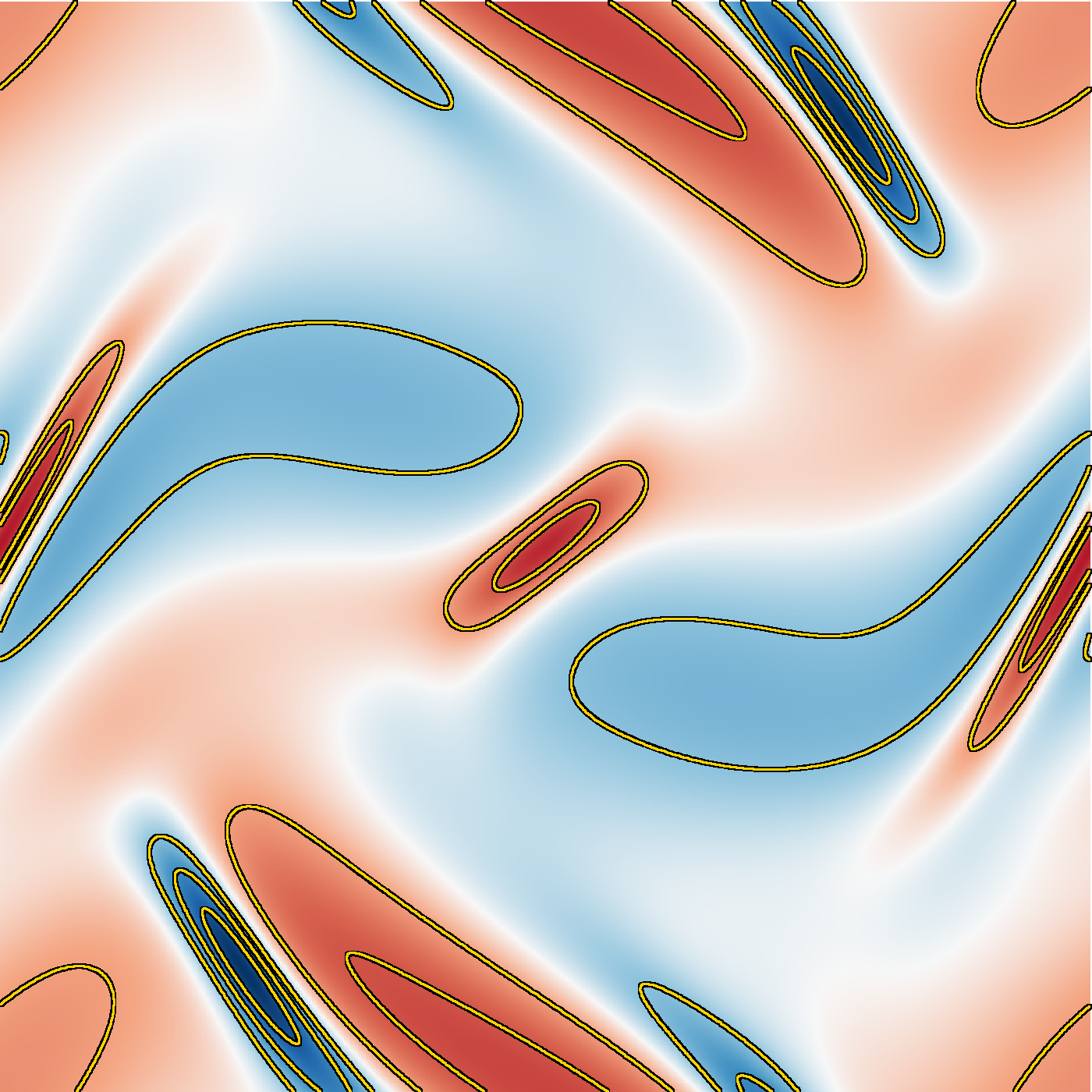};
\nextgroupplot[title={(b) $\mathrm{Re}=200\pi$, $t=1$}, point meta min=-47, point meta max=47]
\addplot graphics[xmin=0, xmax=6.28319, ymin=0, ymax=6.28319] {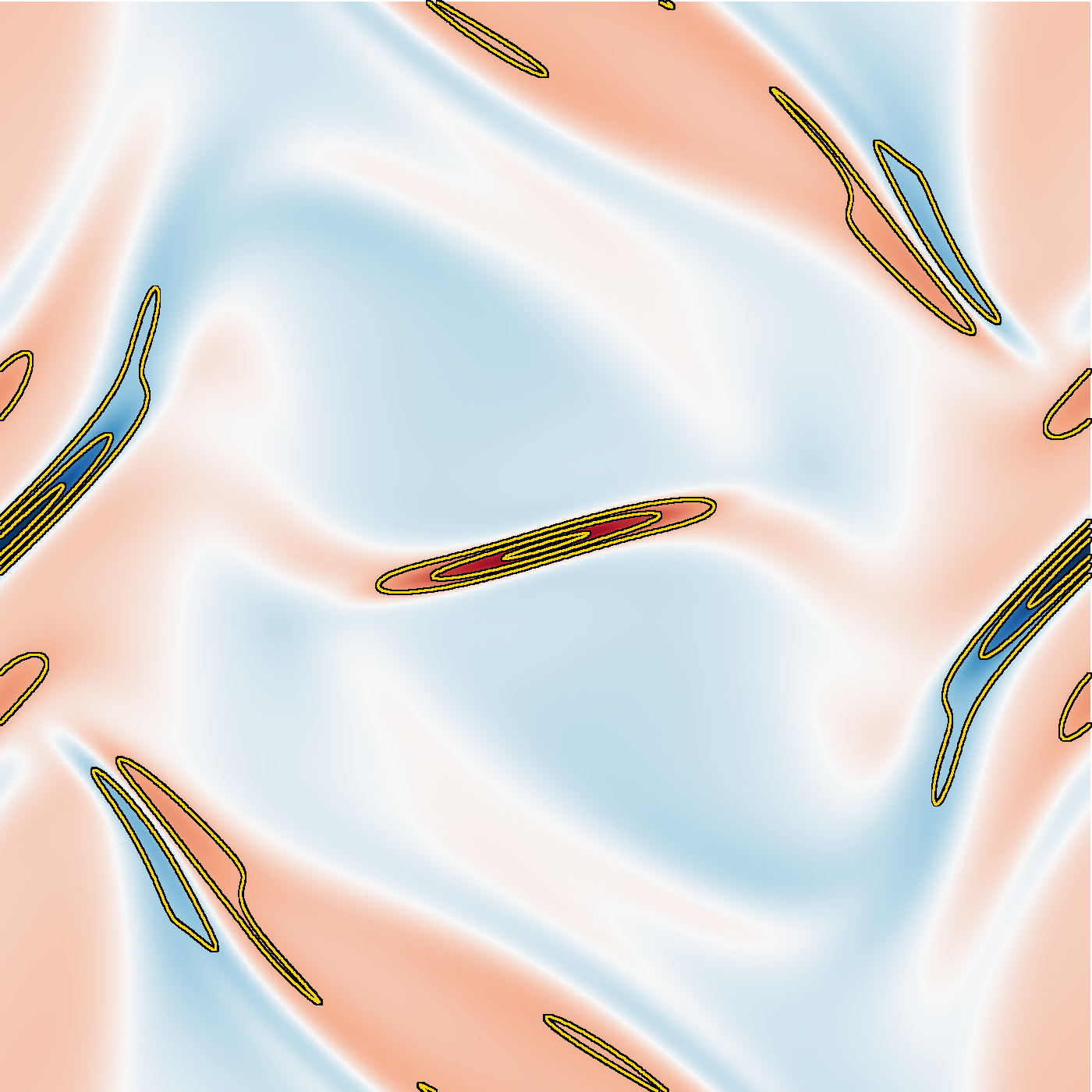};
\nextgroupplot[title={(c) $\mathrm{Re}=5000$, $t=1$}, point meta min=-100, point meta max=100]
\addplot graphics[xmin=0, xmax=6.28319, ymin=0, ymax=6.28319] {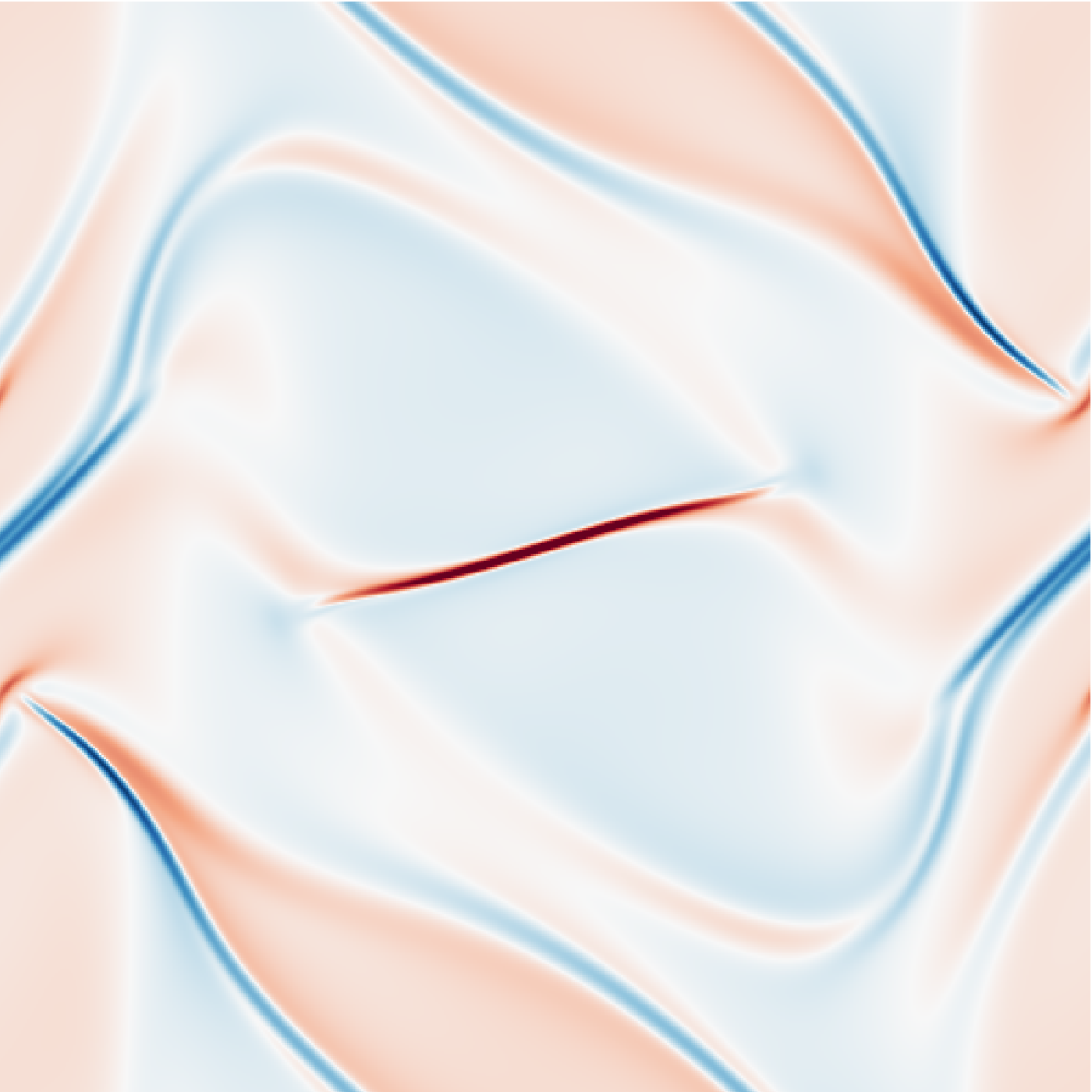};
\nextgroupplot[title={(d) $\mathrm{Re}=5000$, $t=3.5$}, point meta min=-100, point meta max=100]
\addplot graphics[xmin=0, xmax=6.28319, ymin=0, ymax=6.28319] {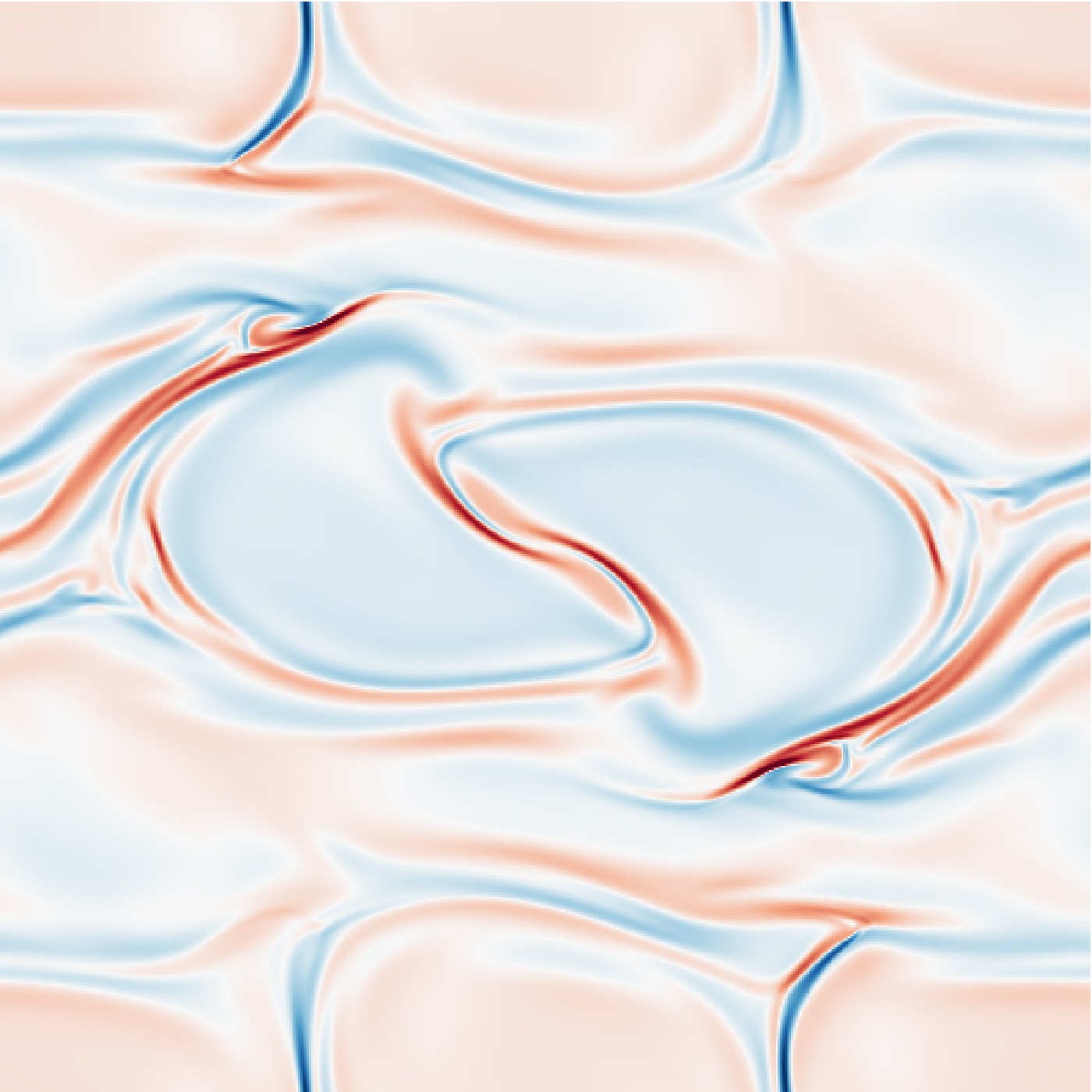};
\end{groupplot}
\end{tikzpicture}
\caption{Current density $j_z$ of the two-dimensional Orszag--Tang vortex computed with RR. Top: $\mathrm{Re}=200\pi$, $N=512$, at $t=0.5$ and $t=1$; the thick black contours are RR and the thin yellow ones, drawn on top, the pseudo-spectral solution ($N_s=512$) at the same levels. Bottom: $\mathrm{Re}=5000$, $N=500$, at $t=1$ and $t=3.5$.}
\label{fig:maps2d}
\end{figure}

RR resolves the current sheets of the vortex (Fig.~\ref{fig:maps2d}); at $\mathrm{Re}=200\pi$ its contours and those of the spectral solution cannot be told apart. The error of $j_{\max}$ (Fig.~\ref{fig:conv2d}) is nearly the same for the four operators and falls at about second order between $N=128$ and $256$ (orders $1.85$--$2.4$). On finer grids the $\mathrm{Ma}^2$ compressibility error of acoustic scaling takes over: halving the lattice velocity at $N=512$ lowers the RR error from $0.073\%$ to $0.026\%$ at $t=0.5$.

\begin{figure}[!htbp]
\centering
\begin{tikzpicture}
\begin{groupplot}[group style={group size=2 by 1, horizontal sep=1.6cm}, width=0.47\linewidth, height=5.6cm,
  xmode=log, ymode=log, log basis x=2, xtick={128,256,512,1024}, xticklabels={128,256,512,1024},
  xmin=100, xmax=1300, ymin=2e-4, ymax=3e-2, xlabel={$N$}, grid=major, grid style={gray!20},
  legend style={draw=none, fill=white, fill opacity=0.85, text opacity=1, font=\footnotesize}, legend cell align=left]
\nextgroupplot[ylabel={$|j_{\max}-j_{\max}^{\rm ref}|/j_{\max}^{\rm ref}$}, title={(a) $t=0.5$}, legend to name=leg:conv2d, legend columns=6, legend style={/tikz/every even column/.append style={column sep=0.3cm}}]
\addplot[cBGK, mark=square*, mark size=1.8pt] table[x=N, y=eBGK05] {data/conv2d.dat};
\addplot[cRM, mark=triangle*, mark size=2.2pt] table[x=N, y=eRM05] {data/conv2d.dat};
\addplot[cCM, mark=diamond*, mark size=2.2pt] table[x=N, y=eCM05] {data/conv2d.dat};
\addplot[cRR, mark=*, mark size=1.6pt] table[x=N, y=eRR05] {data/conv2d.dat};
\addplot[cRR, only marks, mark=o, mark size=3pt, thick] table[x=N, y=eRR05] {data/conv2d_halfMa.dat};
\addplot[black, dashed, domain=128:512, samples=2] {5.5e-3*(128/x)^2};
\legend{BGK, RMs, CMs, RR, {RR, $\tilde u_0/2$}, slope $-2$}
\nextgroupplot[title={(b) $t=1$}]
\addplot[cBGK, mark=square*, mark size=1.8pt] table[x=N, y=eBGK10] {data/conv2d.dat};
\addplot[cRM, mark=triangle*, mark size=2.2pt] table[x=N, y=eRM10] {data/conv2d.dat};
\addplot[cCM, mark=diamond*, mark size=2.2pt] table[x=N, y=eCM10] {data/conv2d.dat};
\addplot[cRR, mark=*, mark size=1.6pt] table[x=N, y=eRR10] {data/conv2d.dat};
\addplot[cRR, only marks, mark=o, mark size=3pt, thick] table[x=N, y=eRR10] {data/conv2d_halfMa.dat};
\addplot[black, dashed, domain=128:512, samples=2] {1.4e-2*(128/x)^2};
\end{groupplot}
\node[anchor=south] at ($(group c1r1.north)!0.5!(group c2r1.north)+(0,0.95cm)$) {\pgfplotslegendfromname{leg:conv2d}};
\end{tikzpicture}
\caption{Relative error of $j_{\max}$ for the two-dimensional vortex at $\mathrm{Re}=200\pi$, against the pseudo-spectral reference ($j_{\max}=18.2631$ at $t=0.5$, $46.7008$ at $t=1$). The open circle is RR with half the lattice velocity, i.e.\ half the Mach number.}
\label{fig:conv2d}
\end{figure}

Section~\ref{sec:equivalence} predicts that the CMs and RMs schemes are $\mathrm{RR}_{\omega,1}$ and $\mathrm{PR}_{\omega,1}$ (Table~\ref{tab:operators}). At $\mathrm{Re}=5000$ and $N=256$ they give the same $j_{\max}$ at every time step up to $t=4$, to within a relative $1.3\times10^{-11}$ and $1.2\times10^{-11}$, i.e.\ round-off amplified by the nonlinear dynamics over $2\times10^4$ time steps. The genuine differences are much larger: the bulk rate, RR against $\mathrm{RR}_{\omega,1}$, changes $j_{\max}$ by up to $1.67\%$, and the filtering frame, $\mathrm{RR}_{\omega,1}$ against $\mathrm{PR}_{\omega,1}$, by up to $1.65\%$. At $N=500$ and $\mathrm{Re}=5000$ (Fig.~\ref{fig:ops2d}) BGK blows up at $t\approx0.36$ ($0.71$ at $\mathrm{Re}=2500$), while the regularised schemes run to $t=4$ within $0.6\%$ of each other. Keeping the aliased components (RR$^*$, a full tensor contraction to fourth order) raises the deviation to $2.75\%$ at the second current peak: the normal stresses are then relaxed at the wrong rate.

\begin{figure}[!htbp]
\centering
\begin{tikzpicture}
\begin{groupplot}[group style={group size=2 by 1, horizontal sep=2.0cm}, width=0.47\linewidth, height=5.6cm,
  xmin=0, xmax=4, xlabel={$t$}, grid=major, grid style={gray!20},
  legend style={draw=none, fill=white, fill opacity=0.85, text opacity=1, font=\footnotesize}, legend cell align=left]
\nextgroupplot[ylabel={$j_{\max}$}, title={(a)}, ymin=0, ymax=230, legend to name=leg:ops2d, legend columns=4, legend style={/tikz/every even column/.append style={column sep=0.4cm}}]
\addplot[cRM, thick] table[x=t, y=RM] {data/ops2d_re5000.dat};
\addplot[cCM, thick, densely dashed] table[x=t, y=CM] {data/ops2d_re5000.dat};
\addplot[cRR, thick, densely dotted] table[x=t, y=RR] {data/ops2d_re5000.dat};
\addplot[cRRs, thick] table[x=t, y=RRstar] {data/ops2d_re5000.dat};
\legend{RMs, CMs, RR, RR$^{*}$ (aliased)}
\nextgroupplot[ylabel={deviation from RMs (\%)}, title={(b)}, ymin=-3, ymax=3,
  every axis plot/.append style={line width=0.45pt}]
\addplot[cCM, thick, densely dashed] table[x=t, y=dCM] {data/ops2d_dev_re5000.dat};
\addplot[cRR, thick, densely dotted] table[x=t, y=dRR] {data/ops2d_dev_re5000.dat};
\addplot[cRRs, thick] table[x=t, y=dRRstar] {data/ops2d_dev_re5000.dat};
\end{groupplot}
\node[anchor=south] at ($(group c1r1.north)!0.5!(group c2r1.north)+(0,0.95cm)$) {\pgfplotslegendfromname{leg:ops2d}};
\end{tikzpicture}
\caption{Two-dimensional vortex at $\mathrm{Re}=5000$, $N=500$: (a) $j_{\max}(t)$; (b) deviation from the RMs scheme. RR$^{*}$ uses a full tensor contraction that also retains the aliased Hermite components. The curves are sampled every $0.002$; the deviations oscillate from sample to sample near the second peak, and the maxima quoted in the text are taken over all samples, every $2\times10^{-4}$.}
\label{fig:ops2d}
\end{figure}

\subsection{Three dimensions: accuracy}\label{sec:res3d}

\begin{figure}[!htb]
\centering
\input{figs/fig_iso3d}
\caption{Isosurfaces of the current density $|\mathbf j|$ of the three-dimensional Orszag--Tang vortex at $\mathrm{Re}=1000$ in the periodic box $[0,2\pi]^3$: RR on $N=128$ (a, c) and the pseudo-spectral solution with $N_s=128$ (b, d), at $t=0.5$ (a, b; $|\mathbf j|=5$) and $t=1$ (c, d; $|\mathbf j|=12$), about $40\%$ of the maximum of the displayed field. For RR, $\mathbf j$ is computed from $\bb$ by spectral differentiation over the wavenumbers kept by the spectral solution, $|k_\alpha|\le42$. Both fields are smoothed with the same Gaussian filter of width $\Delta x$ for display, which removes ripples at the cut-off scale.}
\label{fig:iso3d}
\end{figure}

\begin{figure}[!htbp]
\centering
\begin{tikzpicture}
\begin{groupplot}[group style={group size=2 by 2, horizontal sep=2.4cm, vertical sep=2.2cm},
  scale only axis, width=5.0cm, height=5.0cm, enlargelimits=false, axis on top,
  xmin=0, xmax=6.28319, ymin=0, ymax=6.28319,
  xtick={0,3.14159,6.28319}, xticklabels={$0$,$\pi$,$2\pi$}, ytick={0,3.14159,6.28319}, yticklabels={$0$,$\pi$,$2\pi$},
  xlabel={$x$}, ylabel style={yshift=-0.2cm},
  colorbar, colormap name=ylorrd, colorbar style={width=0.18cm, font=\scriptsize},
  title style={font=\small}]
\nextgroupplot[title={(a) plane $z=0$, $t=0.5$}, ylabel={$y$}, point meta min=0, point meta max=15]
\addplot graphics[xmin=0, xmax=6.28319, ymin=0, ymax=6.28319] {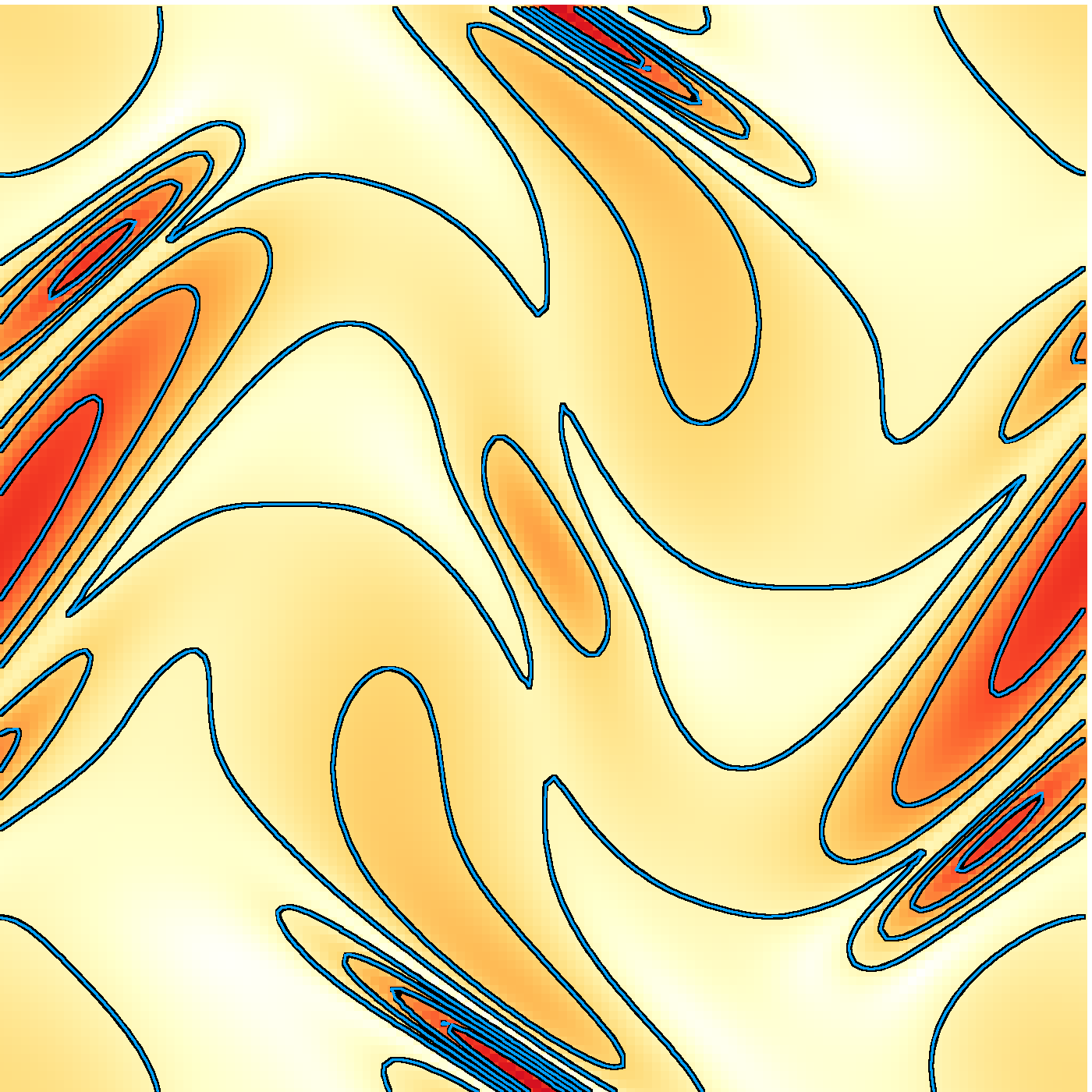};
\nextgroupplot[title={(b) plane $z=0$, $t=1$}, ylabel={$y$}, point meta min=0, point meta max=40]
\addplot graphics[xmin=0, xmax=6.28319, ymin=0, ymax=6.28319] {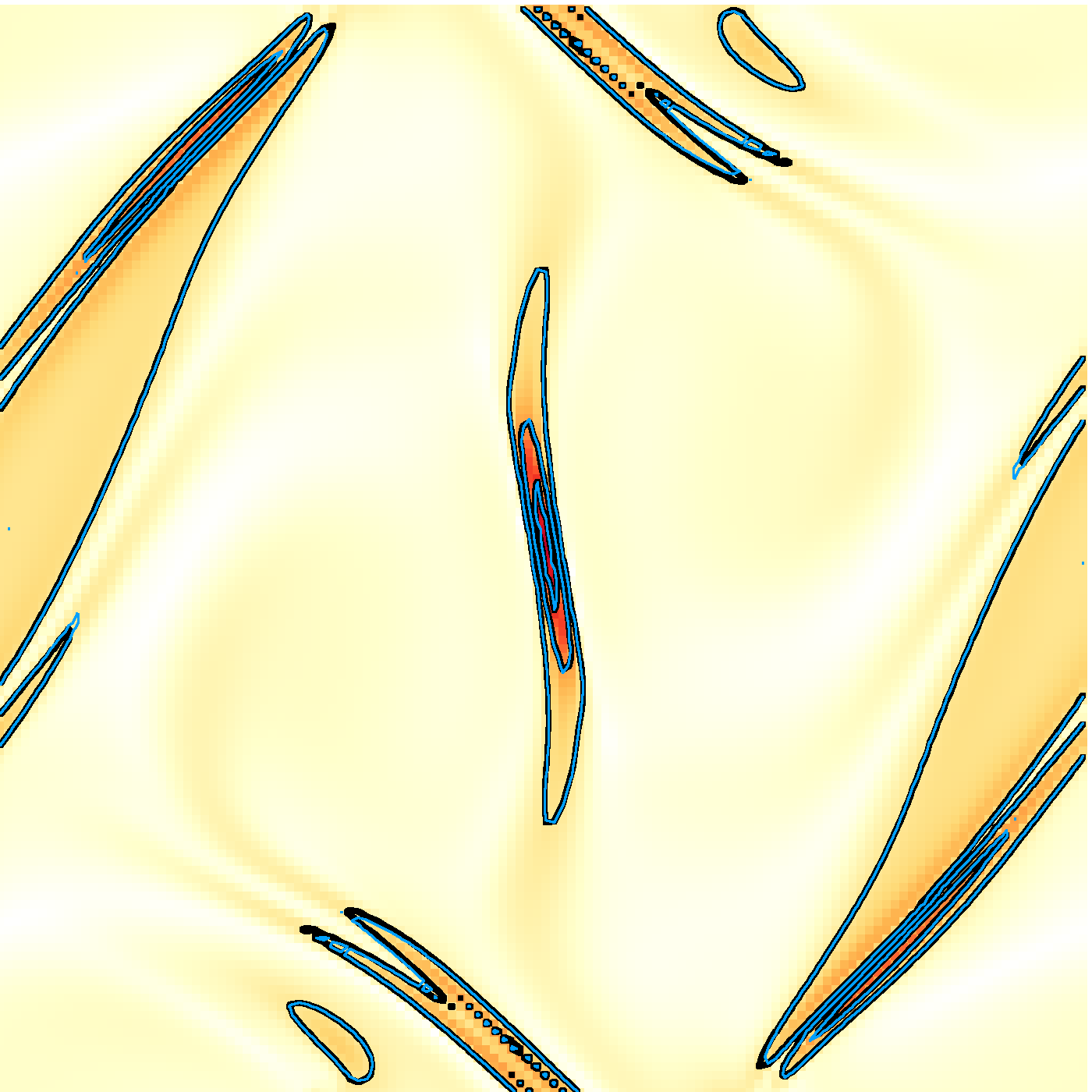};
\nextgroupplot[title={(c) plane $y=0$, $t=0.5$}, ylabel={$z$}, point meta min=0, point meta max=15]
\addplot graphics[xmin=0, xmax=6.28319, ymin=0, ymax=6.28319] {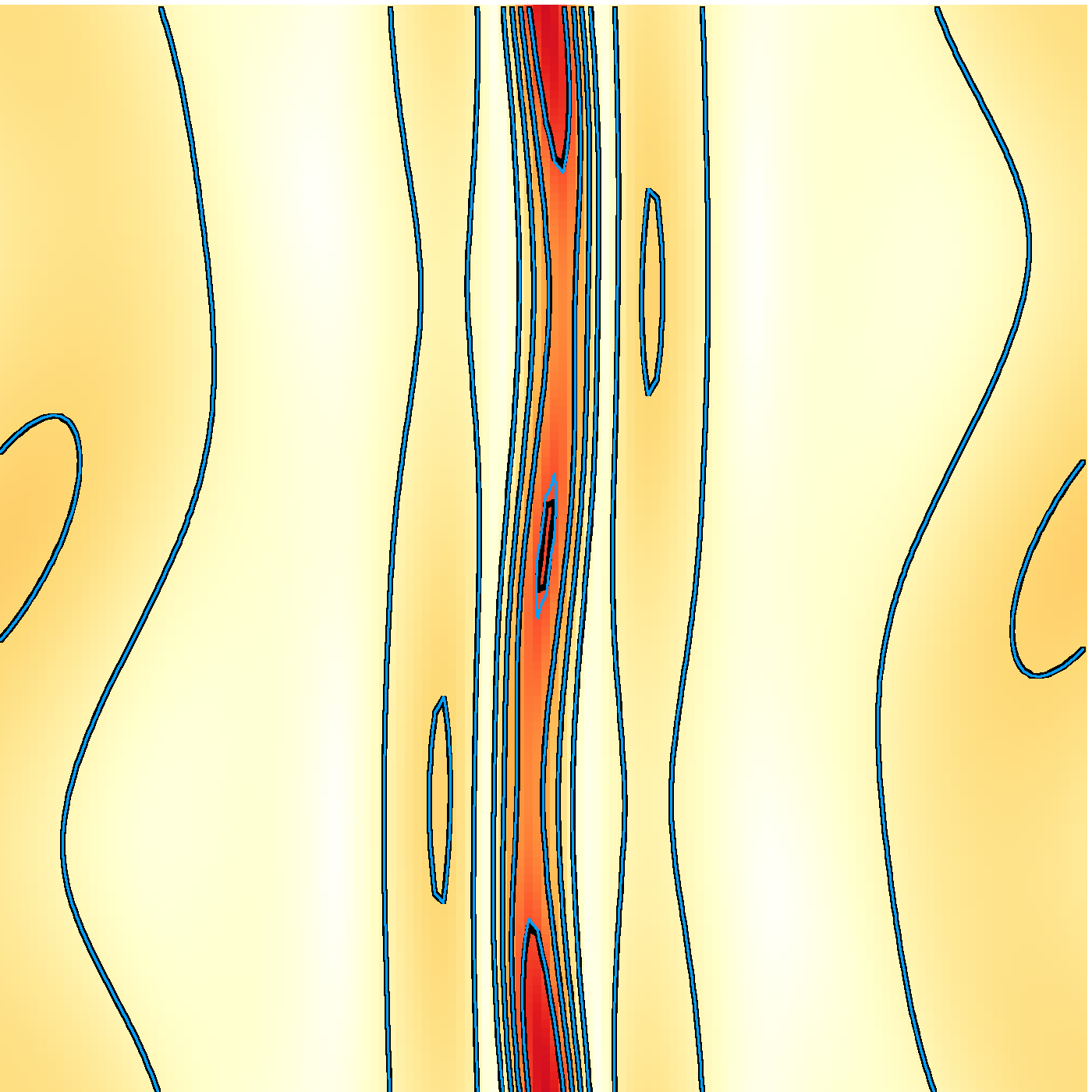};
\nextgroupplot[title={(d) plane $y=0$, $t=1$}, ylabel={$z$}, point meta min=0, point meta max=40]
\addplot graphics[xmin=0, xmax=6.28319, ymin=0, ymax=6.28319] {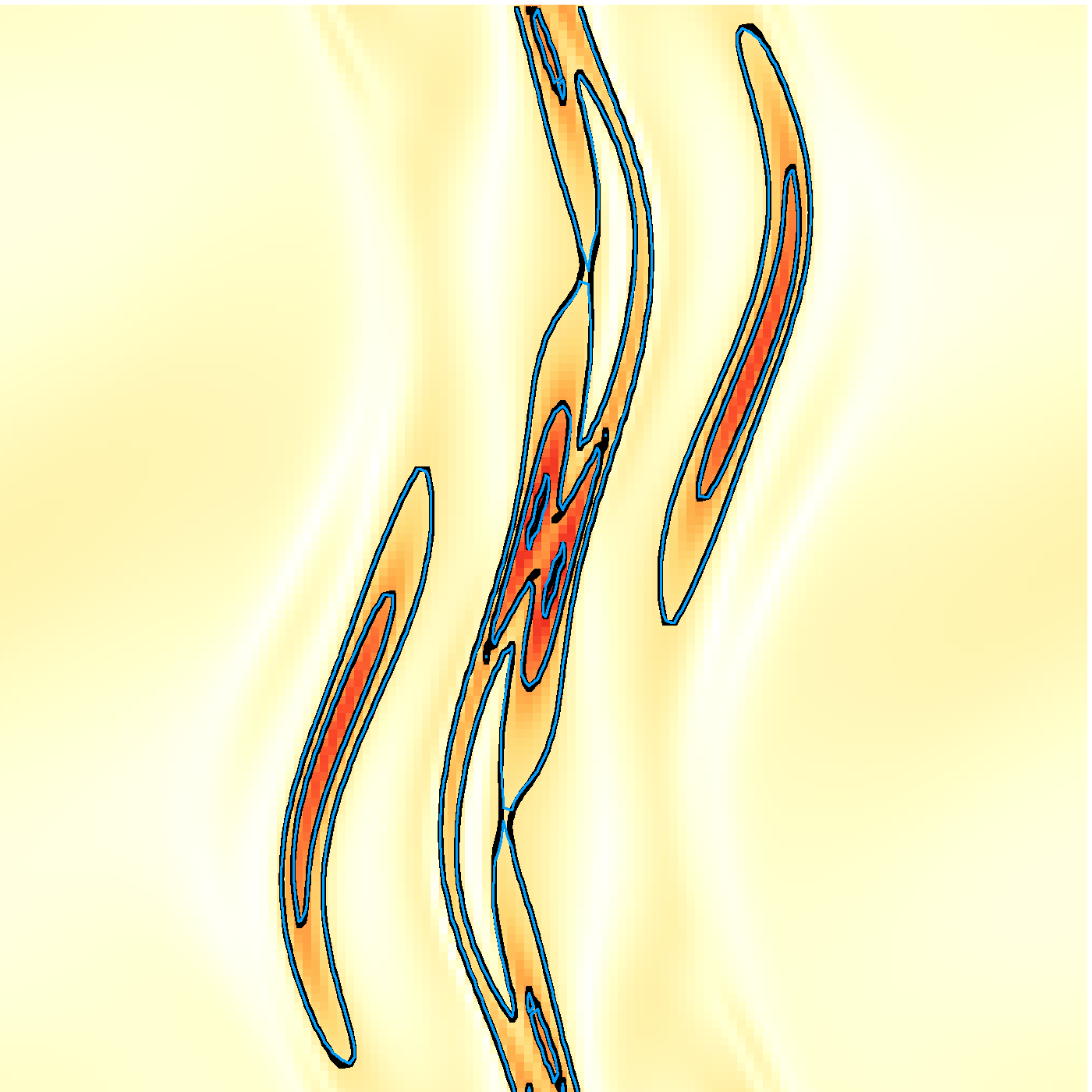};
\end{groupplot}
\end{tikzpicture}
\caption{Current density $|\mathbf j|$ of the three-dimensional Orszag--Tang vortex at $\mathrm{Re}=1000$, computed with RR on $N=128$, on the planes $z=0$ (top) and $y=0$ (bottom) at $t=0.5$ and $t=1$. Thick black contours: RR; thin blue contours, drawn on top: pseudo-spectral solution ($N_s=128$) evaluated with the same central differences on the same nodes; levels 2.5, 5, 7.5, 10 at $t=0.5$ and 10, 20, 30 at $t=1$.}
\label{fig:maps3d}
\end{figure}

At $\mathrm{Re}=1000$ the current sheets of RR on $N=128$ and of the spectral solution are indistinguishable (Figs.~\ref{fig:iso3d} and~\ref{fig:maps3d}); by $t=1$ they have folded and thinned into fully three-dimensional structures a few nodes wide. Over the box the relative $L^2$ difference of the central-difference $|\mathbf j|$ is $0.31\%$ at $t=0.5$ and $3.6\%$ at $t=1$, and the peak current differs by $-1.2\%$ and $-3.0\%$; at $t=1$ the reference itself is marginally resolved, its peak current being $4\%$ below that of $N_s=256$.

At $\mathrm{Re}=100$ the scheme follows the spectral solution closely (Fig.~\ref{fig:hist3d}), and the maximum errors over $0.25\le t\le4$ fall at about second order (Fig.~\ref{fig:conv3d}): $9.2\%$, $0.81\%$, $0.20\%$ and $0.077\%$ in $E_{\rm k}$ and $4.8\%$, $1.2\%$, $0.32\%$ and $0.098\%$ in $E_{\rm m}$ for $N=16$--$128$, and $7.5\%$, $2.7\%$ and $0.67\%$ in $J_{\max}$ for $N=32$--$128$, an order of $1.99$ between the last two grids. The two kinds of diagnostic differ in what they test: the energies are integrals over the box, dominated by the large scales, whereas $J_{\max}$ is the pointwise maximum of a derivative of the field, taken in the thinnest current sheet. Its error is therefore the larger at every resolution, and it is the more demanding of the two. On the finest grid the energy error is partly set by the fixed Mach number: halving the lattice velocity lowers it to $0.048\%$, whereas $J_{\max}$, limited by the grid, keeps its $0.67\%$. BGK is comparable on $N=32$ and $64$ but diverges at $t\approx2$ on $N=16$. The CHM implementation reproduces RR to $3\times10^{-13}$ over 2935 time steps at $N=32$, for $\omega_b=\omega$ and $\omega_b=1$ alike. In this test a monomial central-moment scheme~\cite{derosis2018} with the same equilibrium and $\omega_b=1$, and its higher-order monomial moments at equilibrium, is on the contrary four to twelve times less accurate: $41\%$, $9.9\%$ and $1.5\%$ on $N=16$--$64$, against $0.89\%$ and $0.19\%$ for $\mathrm{RR}_{\omega,1}$ on $N=32$ and $64$. With the equilibrium and the rates the same, the two collisions differ only by the terms~\eqref{eq:monodiff}, through which the monomial one leaves part of the relaxed stress, deviatoric components included, in the fourth-order Hermite moments: this is what costs the accuracy. The comparison concerns this collision and this flow; it is not a statement about monomial moments as such, which reproduce RR if their higher-order moments are set as described in Section~\ref{sec:equivalence}. Truncating the equilibrium at second order, which in our implementation also truncates the reconstruction~\eqref{eq:closed} at second order and so turns RR into projective regularisation, costs almost nothing ($9.1\%$, $0.79\%$ and $0.19\%$, open circles in Fig.~\ref{fig:conv3d}).

\begin{figure}[!htbp]
\centering
\begin{tikzpicture}
\begin{groupplot}[group style={group size=3 by 1, horizontal sep=1.45cm}, width=0.32\linewidth, height=5.0cm,
  xmin=0, xmax=4, xlabel={$t$}, grid=major, grid style={gray!20}, every axis title/.style={font=\small, at={(0.5,1)}, above},
  legend style={draw=none, fill=white, fill opacity=0.85, text opacity=1, font=\scriptsize}, legend cell align=left]
\nextgroupplot[title={(a) $E_{\rm k}/E_{\rm k}(0)$}, ymin=0, ymax=1.05, legend to name=leg:hist3d, legend columns=4, legend style={font=\footnotesize, /tikz/every even column/.append style={column sep=0.4cm}}]
\addplot[black, thick] table[x=t, y=Gamma] {data/spec3d.dat};
\addplot[cN32, densely dashed] table[x=t, y=Gamma] {data/hist3d_N32.dat};
\addplot[cN64, densely dashed] table[x=t, y=Gamma] {data/hist3d_N64.dat};
\addplot[cN128, densely dotted, thick] table[x=t, y=Gamma] {data/hist3d_N128.dat};
\legend{spectral, $N=32$, $N=64$, $N=128$}
\nextgroupplot[title={(b) $E_{\rm m}$}, ymin=0, ymax=2.1]
\addplot[black, thick] table[x=t, y=Em] {data/spec3d.dat};
\addplot[cN32, densely dashed] table[x=t, y=Em] {data/hist3d_N32.dat};
\addplot[cN64, densely dashed] table[x=t, y=Em] {data/hist3d_N64.dat};
\addplot[cN128, densely dotted, thick] table[x=t, y=Em] {data/hist3d_N128.dat};
\nextgroupplot[title={(c) $J_{\max}$}, ymin=0, ymax=22]
\addplot[cN32!45, thick] table[x=t, y=Jfd32] {data/spec3d.dat};
\addplot[cN64!45, thick] table[x=t, y=Jfd64] {data/spec3d.dat};
\addplot[cN128!45, thick] table[x=t, y=Jfd128] {data/spec3d.dat};
\addplot[cN32, densely dashed] table[x=t, y=Jfd] {data/hist3d_N32.dat};
\addplot[cN64, densely dashed] table[x=t, y=Jfd] {data/hist3d_N64.dat};
\addplot[cN128, densely dotted, thick] table[x=t, y=Jfd] {data/hist3d_N128.dat};
\end{groupplot}
\node[anchor=south] at ($(group c1r1.north)!0.5!(group c3r1.north)+(0,0.95cm)$) {\pgfplotslegendfromname{leg:hist3d}};
\end{tikzpicture}
\caption{Three-dimensional vortex at $\mathrm{Re}=100$: kinetic energy, magnetic energy and peak current of the D3Q27 RR scheme against the pseudo-spectral solution. In (c), the light lines are the spectral solution evaluated with the lattice stencil at $N=32$, $64$ and $128$.}
\label{fig:hist3d}
\end{figure}

\begin{figure}[!htbp]
\centering
\begin{tikzpicture}
\begin{loglogaxis}[width=0.5\linewidth, height=6.0cm, log basis x=2, xtick={16,32,64,128}, xticklabels={16,32,64,128},
  xmin=13, xmax=150, ymin=3e-4, ymax=5, xlabel={$N$}, ylabel={maximum relative error},
  grid=major, grid style={gray!20}, legend style={at={(1.03,1)}, anchor=north west, draw=none, font=\footnotesize}, legend cell align=left]
\addplot[cRR, mark=*, thick] table[x=N, y=eG] {data/conv3d.dat};
\addplot[cRR, mark=square, thick, densely dashed] table[x=N, y=eEm] {data/conv3d.dat};
\addplot[cRR, mark=triangle, thick, densely dotted] table[x=N, y=eJ] {data/conv3d.dat};
\addplot[cBGK, mark=x, mark size=3pt, thick] table[x=N, y=eG_BGK] {data/conv3d.dat};
\addplot[cCM, mark=diamond*, thick] table[x=N, y=eG_CMmono] {data/conv3d.dat};
\addplot[cCHM, mark=pentagon*, thick] table[x=N, y=eG_CMS] {data/conv3d.dat};
\addplot[cRR, only marks, mark=o, mark size=3.2pt] table[x=N, y=eG_RReq2] {data/conv3d.dat};
\addplot[black, dashed, domain=32:128, samples=2] {0.011*(32/x)^2};
\legend{RR: $E_{\rm k}$, RR: $E_{\rm m}$, RR: $J_{\max}$, BGK: $E_{\rm k}$, {CM (monomial, $\omega_b=1$): $E_{\rm k}$}, {$\mathrm{RR}_{\omega,1}$ (= CHM): $E_{\rm k}$}, {PR, second-order $f^{\rm eq}$: $E_{\rm k}$}, slope $-2$}
\end{loglogaxis}
\end{tikzpicture}
\caption{Maximum relative errors over $0.25\le t\le4$ for the three-dimensional vortex at $\mathrm{Re}=100$, against the pseudo-spectral solution.}
\label{fig:conv3d}
\end{figure}

An oblique shear Alfv\'en wave confirms second order for the coupled scheme. With a uniform field $0.02(1,1,1)/\sqrt3$ in lattice units, wave vector $(1,1,1)2\pi/L$, velocity and field perturbations along $(1,-1,0)/\sqrt2$ with velocity amplitude $0.005$, and $\nu=\eta=0.01$, phase speed and damping rate are fitted over 1500 steps to the projections of $\bu$ and $\bb$ on the cosine and sine of the phase. The phase error is $-1.41\%$, $-0.35\%$ and $-0.087\%$ for $N=16$, $32$ and $64$, the damping-rate error $+0.98\%$, $+0.17\%$ and $+0.020\%$, and both are equally small for $\mathrm{Pr_m}=0.3$ and $3$.

A uniform drift along $x$, of lattice velocity $U_0$ ($N=32$, $\mathrm{Re}=100$, $t\le4$; Fig.~\ref{fig:galilean}), separates three Galilean defects. The shear stress is the first. Without a field, the MHD vortex is a decaying Taylor--Green flow with pure shear strain, and with the complete or the third-order equilibrium (the reconstruction truncated at the same order) its effective viscosity, fitted to $\ln E_{\rm k}$ over $0.5\le t\le4$, changes by less than $0.15$ points up to $U_0=0.2$ ($\mathrm{Ma}=0.35$). With the second-order equilibrium RR is projective regularisation, which is linearly unstable above $U_0\approx0.095$ (\ref{app:lsa}): its viscosity error reaches $-1.9\%$ at $U_0=0.1$, and it diverges at $U_0=0.2$, where the unstable grid-scale mode grows by a factor $e^{39}$ in the 830 steps before the divergence. The normal stresses are the second. Since $c^3=c$ on D1Q3, the product-form equilibrium cannot carry $\rho u_\alpha^3$; a three-dimensional Taylor--Green vortex, $\bu=(\sin x\cos y\cos z,-\cos x\sin y\cos z,0)$ in physical units, whose strain is not a pure shear, gains $0.62\%$, $2.5\%$ and $10.4\%$ of kinetic energy at $t=4$ for $U_0=0.05$, $0.1$ and $0.2$, the cubic defect of the lattice~\cite{dellar2014}. The magnetic populations are the third: their first-order equilibrium makes the diffusivity drift dependent, and a weak passive field gains $0.84\%$, $3.2\%$ and $12.6\%$ of energy. To isolate the fluid side of full MHD, we prescribe the field as the initial field advected with the drift. The kinetic energy then changes by $0.23\%$, $0.85\%$ and $2.4\%$, and the comoving Maxwell term changes these numbers by $0.007$, $0.024$ and $0.08$ points; down to $U_0=0.01$, below the peak field of $0.025$ in lattice units, both grow roughly like $U_0^2$, the difference staying near $3\%$ of the change. The frame dependence~\eqref{eq:L} of Dellar's Maxwell term acts through the local velocity, so it grows with the field: when a field three times stronger drives the flow to $|\bu|\approx0.2$, the two Maxwell terms differ by $23\%$ in $E_{\rm k}$ without any drift. With the standard field, however, the hierarchy of the Galilean defects is clear. The second-order equilibrium makes the shear stress drift dependent, and the scheme unstable beyond $U_0\approx0.095$; the complete equilibrium removes this defect, to $0.15$ points. What remains at $U_0=0.1$ is the drift-dependent magnetic diffusivity ($3.2\%$) and the cubic defect of the normal stresses ($2.5\%$ in the Taylor--Green vortex, $0.85\%$ in the vortex with a prescribed field), of which the frame dependence of the Maxwell term accounts for $0.024$ points. The Maxwell term is therefore not the dominant source of Galilean error in the standard configuration.

\begin{figure}[!htbp]
\centering
\begin{tikzpicture}
\begin{groupplot}[group style={group size=2 by 1, horizontal sep=2.1cm}, width=0.47\linewidth, height=5.6cm, grid=major, grid style={gray!20},
  legend style={draw=none, fill=white, fill opacity=0.85, text opacity=1, font=\footnotesize}, legend cell align=left]
\nextgroupplot[xlabel={$U_0$ (lattice units)}, ylabel={$\nu_{\rm eff}/\nu-1$ (\%)}, title={(a) fluid, $\mathbf b=\mathbf 0$}, xmin=-0.01, xmax=0.21, xtick={0,0.05,0.1,0.15,0.2}, xticklabel style={/pgf/number format/fixed, /pgf/number format/precision=2}, ymin=-6.5, ymax=1.5, legend to name=leg:gal, legend columns=3, legend style={/tikz/every even column/.append style={column sep=0.4cm}}]
\addplot[cRR, mark=*, thick] table[x=U0, y expr=100*\thisrow{RR6}] {data/galilean_fluid.dat};
\addplot[cRRL, mark=o, thick, densely dashed] table[x=U0, y expr=100*\thisrow{RR3}] {data/galilean_fluid.dat};
\addplot[cPR, mark=triangle*, thick] table[x=U0, y expr=100*\thisrow{RR2}] {data/galilean_fluid.dat};
\addplot[cBGK, mark=square*, thick] table[x=U0, y expr=100*\thisrow{BGK6}] {data/galilean_fluid.dat};
\addplot[cBGK!50, mark=square, thick, densely dashed] table[x=U0, y expr=100*\thisrow{BGK2}] {data/galilean_fluid.dat};
\legend{{RR, 6th-order $f^{\rm eq}$}, {RR, 3rd order}, {RR, 2nd order}, {BGK, 6th order}, {BGK, 2nd order}}
\nextgroupplot[xmode=log, ymode=log, xlabel={$U_0$ (lattice units)}, ylabel={relative change at $t=4$}, title={(b) field effects and normal strain}, xmin=0.008, xmax=0.25, ymin=5e-5, ymax=0.6, xtick={0.01,0.02,0.05,0.1,0.2}, xticklabels={$0.01$,$0.02$,$0.05$,$0.1$,$0.2$}, legend to name=leg:galb, legend columns=3, legend style={font=\scriptsize, /tikz/every even column/.append style={column sep=0.3cm}}]
\addplot[cCM, mark=*, thick] table[x=U0, y=dEm_weak] {data/galilean_mag.dat};
\addplot[cRR, mark=square*, thick] table[x=U0, y=dG_std] {data/galilean_mag.dat};
\addplot[cRRL, mark=o, thick, densely dashed] table[x=U0, y=dG_cov] {data/galilean_mag.dat};
\addplot[cRM, mark=diamond*, thick] table[x=U0, y=dG_kin_dellar] {data/galilean_kin.dat};
\addplot[cRM!60, mark=diamond, thick, densely dashed] table[x=U0, y=dG_kin_cov] {data/galilean_kin.dat};
\addplot[cBGK, mark=triangle*, thick] table[x=U0, y=dG_tgv] {data/galilean_kin.dat};
\addplot[black, dashed] table[x=U0, y=U2_over_cs2m] {data/galilean_mag.dat};
\legend{{$E_{\rm m}$, passive field}, {$E_{\rm k}$, MHD, Dellar}, {$E_{\rm k}$, MHD, comoving}, {$E_{\rm k}$, prescribed field, Dellar}, {$E_{\rm k}$, prescribed field, comoving}, {$E_{\rm k}$, Taylor--Green, $\mathbf b=\mathbf 0$}, $U_0^2/c_{s,\rm m}^2$}
\end{groupplot}
\node[anchor=south] at ($(group c1r1.north)!0.5!(group c2r1.north)+(0,0.95cm)$) {\pgfplotslegendfromname{leg:gal}};
\node[anchor=north] at ($(group c1r1.south)!0.5!(group c2r1.south)+(0,-1.15cm)$) {\pgfplotslegendfromname{leg:galb}};
\end{tikzpicture}
\caption{Uniform drift $U_0$ at $N=32$. (a) Error of the effective viscosity of the decaying Taylor--Green flow of the Orszag--Tang velocity ($\bb=\mathbf 0$, pure shear) for RR and BGK with complete, third- and second-order equilibria. (b) Change at $t=4$, relative to the run without drift, of the magnetic energy of a weak passive field, of the kinetic energy in full MHD and with a prescribed field advected with the drift, each with Dellar's and with the comoving Maxwell term~\eqref{eq:comoving}, and of the kinetic energy of the three-dimensional Taylor--Green vortex ($\bb=\mathbf 0$); RR with the complete equilibrium.}
\label{fig:galilean}
\end{figure}

At $\mathrm{Re}=1000$ the current peaks at $t\approx2.1$ (Fig.~\ref{fig:re1000}), and the spectral reference with $N_s=256$ is resolved up to $t=3$. RR follows it through the peak: over $0.25\le t\le3$ the maximum errors in $E_{\rm k}$, $E_{\rm m}$ and $J_{\max}$ are $0.27\%$, $0.12\%$ and $9.3\%$ on $N=128$, and $0.090\%$, $0.063\%$ and $4.0\%$ on $N=256$, while BGK diverges at $t\approx0.70$. Between the two grids the errors fall by factors of $3.0$, $1.9$ and $2.3$, short of the factor four of second order: with current sheets a few nodes wide these runs are not in the asymptotic range, and the comparison at this Reynolds number shows agreement with the reference, not the order of convergence, which is measured at $\mathrm{Re}=100$. The error of $J_{\max}$ exceeds those of the energies by more than an order of magnitude for the reason given above. The spectra (Fig.~\ref{fig:spec3d}) locate the remaining error. At $t\approx2$ the kinetic spectrum is within $2\%$ of the reference up to $k\Delta x\approx0.5$, about 13 nodes per wavelength, and within $10\%$ up to $0.8$, on both grids, and more dissipative beyond. The magnetic spectrum is within $2\%$ up to $k\Delta x\approx0.75$; beyond, its ratio does not collapse in $k\Delta x$ ($0.98$ on $N=128$ against $0.86$ on $N=256$ at $k\Delta x=1.5$), consistent with grid-scale noise of the magnetic populations, the node-to-node ripples of Fig.~\ref{fig:maps3d}, adding energy at high $k$ on the coarser grid.

\begin{figure}[!htbp]
\centering
\begin{tikzpicture}
\begin{groupplot}[group style={group size=2 by 1, horizontal sep=1.6cm}, width=0.47\linewidth, height=5.4cm, xmin=0, xmax=3, xlabel={$t$}, grid=major, grid style={gray!20}, legend style={draw=none, fill=white, fill opacity=0.85, text opacity=1, font=\footnotesize}, legend cell align=left]
\nextgroupplot[title={(a) $E_{\rm k}/E_{\rm k}(0)$}, ymin=0.4, ymax=1.03, legend to name=leg:re1000, legend columns=4, legend style={/tikz/every even column/.append style={column sep=0.4cm}}]
\addplot[black, thick] table[x=t, y=Gamma] {data/spec3d_re1000_ns256.dat};
\addplot[cCM, thick, dashdotted] table[x=t, y=Gamma] {data/gpu_re1000_m128_cm_wbomega.dat};
\addplot[cRR, thick, densely dashed] table[x=t, y=Gamma] {data/gpu_re1000_m256_cm_wbomega.dat};
\addplot[cBGK, thick, densely dotted] table[x=t, y=Gamma] {data/gpu_re1000_m128_bgk.dat};
\legend{spectral ($N_s=256$), {RR, $N=128$}, {RR, $N=256$}, {BGK, $N=128$}}
\nextgroupplot[title={(b) $J_{\max}$}, ymin=0, ymax=115]
\addplot[gray!45, line width=2.6pt] table[x=t, y=Jfd128] {data/spec3d_re1000_ns256.dat};
\addplot[black, thick] table[x=t, y=Jfd256] {data/spec3d_re1000_ns256.dat};
\addplot[cCM, thick, dashdotted] table[x=t, y=J] {data/gpu_re1000_m128_cm_wbomega.dat};
\addplot[cRR, thick, densely dashed] table[x=t, y=J] {data/gpu_re1000_m256_cm_wbomega.dat};
\addplot[cBGK, thick, densely dotted] table[x=t, y=J] {data/gpu_re1000_m128_bgk.dat};
\addplot[cBGK, only marks, mark=x, mark size=3pt, thick] coordinates {(0.5969,31.5939)};
\end{groupplot}
\node[anchor=south] at ($(group c1r1.north)!0.5!(group c2r1.north)+(0,0.95cm)$) {\pgfplotslegendfromname{leg:re1000}};
\end{tikzpicture}
\caption{Three-dimensional vortex at $\mathrm{Re}=1000$: RR on $N=128$ and $N=256$ and BGK on $N=128$ (cross: last sample before the blow-up), against the pseudo-spectral solution ($N_s=256$). In (b) the spectral solution is evaluated with the lattice stencil at $N=256$ (black) and $N=128$ (wide grey line).}
\label{fig:re1000}
\end{figure}

\begin{figure}[!htbp]
\centering
\begin{tikzpicture}
\begin{groupplot}[group style={group size=2 by 1, horizontal sep=1.8cm}, width=0.47\linewidth, height=5.8cm, grid=major, grid style={gray!20}, legend style={draw=none, fill=white, fill opacity=0.85, text opacity=1, font=\footnotesize}, legend cell align=left]
\nextgroupplot[xmode=log, ymode=log, xmin=1, xmax=300, ymin=1e-10, ymax=3, xlabel={$k$}, ylabel={$E(k)$}, title={(a) spectra at $t\approx2$}, legend to name=leg:spec, legend columns=3, legend style={/tikz/every even column/.append style={column sep=0.35cm}}]
\addplot[black, thick] table[x=k, y=Eu_ref] {data/spec_re1000_m256_f4.dat};
\addplot[cRR, thick, densely dashed] table[x=k, y=Eu] {data/spec_re1000_m256_f4.dat};
\addplot[cPR, thick] table[x=k, y=Eu] {data/spec_re3000_m512_wb1_f4.dat};
\legend{{spectral, $\mathrm{Re}=1000$}, {RR, $N=256$, $\mathrm{Re}=1000$}, {$\mathrm{RR}_{\omega,1}$, $N=512$, $\mathrm{Re}=3000$}}
\addplot[black, thin, forget plot] table[x=k, y=Eb_ref] {data/spec_re1000_m256_f4.dat};
\addplot[cRR, thin, densely dashed, forget plot] table[x=k, y=Eb] {data/spec_re1000_m256_f4.dat};
\addplot[cPR, thin, forget plot] table[x=k, y=Eb] {data/spec_re3000_m512_wb1_f4.dat};
\nextgroupplot[xmin=0, xmax=2, ymin=0, ymax=1.1, xlabel={$k\Delta x$}, ylabel={$E^{\rm LB}(k)/E^{\rm ref}(k)$}, title={(b) $\mathrm{Re}=1000$, $t\approx2$}, legend pos=south west, legend style={font=\scriptsize}]
\addplot[gray!15, fill=gray!15, forget plot] coordinates {(0,0.98) (2,0.98) (2,1.02) (0,1.02)} -- cycle;
\addplot[cCM, thick] table[x=kdx, y=rEu] {data/spec_re1000_m128_f4.dat};
\addplot[cRR, thick] table[x=kdx, y=rEu] {data/spec_re1000_m256_f4.dat};
\addplot[cCM, thick, densely dashed] table[x=kdx, y=rEb] {data/spec_re1000_m128_f4.dat};
\addplot[cRR, thick, densely dashed] table[x=kdx, y=rEb] {data/spec_re1000_m256_f4.dat};
\legend{{$E_{\rm u}$, $N=128$}, {$E_{\rm u}$, $N=256$}, {$E_{\rm b}$, $N=128$}, {$E_{\rm b}$, $N=256$}}
\end{groupplot}
\node[anchor=south] at ($(group c1r1.north)!0.5!(group c2r1.north)+(0,0.95cm)$) {\pgfplotslegendfromname{leg:spec}};
\end{tikzpicture}
\caption{Energy spectra at $t\approx2$. (a) Kinetic (thick) and magnetic (thin) spectra at $\mathrm{Re}=1000$, pseudo-spectral and RR on $N=256$, and $\mathrm{RR}_{\omega,1}$ at $\mathrm{Re}=3000$ on $N=512$. (b) Ratio of the RR spectra to the pseudo-spectral ones at $\mathrm{Re}=1000$, against $k\Delta x$, on $N=128$ and $N=256$; the grey band is $\pm2\%$.}
\label{fig:spec3d}
\end{figure}

\subsection{Three dimensions: robustness}\label{sec:reshigh}

At higher Reynolds numbers the outcome depends on the collision and on the bulk rate (Table~\ref{tab:div}, Fig.~\ref{fig:highre}). BGK diverges first, the earlier the finer the grid, at $t\approx0.25$--$0.75$ at $\mathrm{Re}=3000$ and at $0.15$--$0.30$ at $\mathrm{Re}=10\,000$. Single-rate RR lasts more than twice as long but fails before the current peak, and at the fixed lattice velocity of acoustic scaling it fails earlier on finer grids, at $t\approx1.7$ on $N=64$ and $128$ but $0.85$ and $0.65$ on $N=256$ and $512$: the failure is local, not a lack of resolution. With the stress trace relaxed to equilibrium, $\omega_b=1$, RR runs through the peak to $t=3$ on $N=256$ and $512$ and follows the pseudo-spectral solution at this Reynolds number ($N_s=512$, \ref{app:spectral}; Fig.~\ref{fig:highre}a): over $0.25\le t\le3$ its errors in $E_{\rm k}$, $E_{\rm m}$ and $J_{\max}$ are $0.24\%$, $0.063\%$ and $16\%$ on $N=256$, and $0.049\%$, $0.050\%$ and $7.4\%$ on $N=512$. The peak current, $222$ and $296$ at $t\approx2.3$, is within $1\%$ of the reference evaluated with the same stencil on the same nodes ($222$ and $295$): the two grids differ through the stencil, which returns $35\%$ and $14\%$ less than the spectral peak of $343$, not through the solution. On $N=128$ it reaches $t=3$ as well, but on $N=64$ the unit bulk rate only delays the failure, from $t\approx1.66$ to $1.71$. Where both rates run, the unit bulk rate costs nothing: at $\mathrm{Re}=1000$ on $N=256$, $\mathrm{RR}_{\omega,1}$ and single-rate RR differ by at most $0.1\%$, and the errors of $\mathrm{RR}_{\omega,1}$ in $E_{\rm k}$, $E_{\rm m}$ and $J_{\max}$ are $0.049\%$, $0.060\%$ and $4.0\%$. The bulk viscosity $\nu_b=\tfrac23\cs(1/\omega_b-\tfrac12)$, $1/9$ in lattice units for $\omega_b=1$, is $94$ times the shear viscosity of the $N=512$ run; bulk viscosity is known to stabilise hydrodynamic lattice Boltzmann schemes~\cite{dellar2001}, and it is also the rate of the two-dimensional CMs and RMs schemes. It overdamps sound waves as well~\cite{coreixas2019}: the remedy suits the low-Mach-number, nearly incompressible flows considered here, and its use where acoustics or compressibility matter would need a separate study. At $\mathrm{Re}=10\,000$ it only delays the failure, from $t\approx0.65$ to $1.10$ on $N=256$ and from $0.35$ to $0.90$ on $N=512$.

\begin{table}[!htbp]
\centering\footnotesize
\caption{Divergence of the three-dimensional vortex: time of the first sample with non-finite fields or a velocity beyond the lattice speed (samples every $0.01$ on $N=64$ and on $N=128$ at $\mathrm{Re}=3000$, every $0.05$ otherwise). --: runs to the end ($t=3$; $t=2$ on $N=64$ at $\mathrm{Re}=1000$).}
\label{tab:div}
\begin{tabular}{@{}lccccccccc@{}}
\toprule
 & \multicolumn{3}{c}{$\mathrm{Re}=1000$} & \multicolumn{4}{c}{$\mathrm{Re}=3000$} & \multicolumn{2}{c}{$\mathrm{Re}=10\,000$}\\
\cmidrule(lr){2-4}\cmidrule(lr){5-8}\cmidrule(l){9-10}
$N$ & 64 & 128 & 256 & 64 & 128 & 256 & 512 & 256 & 512\\
\midrule
BGK & 0.79 & 0.70 & 0.55 & 0.75 & 0.51 & 0.35 & 0.25 & 0.30 & 0.15\\
RR & -- & -- & -- & 1.66 & 1.78 & 0.85 & 0.65 & 0.65 & 0.35\\
$\mathrm{RR}_{\omega,1}$ & -- & -- & -- & 1.71 & -- & -- & -- & 1.10 & 0.90\\
\bottomrule
\end{tabular}
\end{table}

\begin{figure}[!htbp]
\centering
\begin{tikzpicture}
\begin{groupplot}[group style={group size=2 by 1, horizontal sep=1.8cm}, width=0.47\linewidth, height=5.6cm, xmin=0, xmax=3, xlabel={$t$}, grid=major, grid style={gray!20}, legend style={draw=none, fill=white, fill opacity=0.85, text opacity=1, font=\footnotesize}, legend cell align=left]
\nextgroupplot[title={(a) $J_{\max}$}, ymin=0, ymax=320, legend to name=leg:highre, legend columns=3, legend style={/tikz/every even column/.append style={column sep=0.4cm}}]
\addplot[gray!55, line width=2.6pt] table[x=t, y=Jfd512] {data/spec3d_re3000_ns512.dat};
\addplot[gray!30, line width=2.6pt, forget plot] table[x=t, y=Jfd256] {data/spec3d_re3000_ns512.dat};
\addplot[cRR, thick] table[x=t, y=J] {data/gpu_re3000_m512_cm_wb1.dat};
\addplot[cRRL, thick] table[x=t, y=J] {data/gpu_re3000_m256_cm_wb1.dat};
\addplot[cRR, thick, densely dashed] table[x=t, y=J] {data/gpu_re3000_m512_cm_wbomega.dat};
\addplot[cCM, thick, dashdotted] table[x=t, y=J] {data/gpu_re3000_m256_cm_wbomega.dat};
\addplot[cBGK, thick, densely dotted] table[x=t, y=J] {data/gpu_re3000_m512_bgk.dat};
\addplot[cBGK!50, thick, densely dotted] table[x=t, y=J] {data/gpu_re3000_m256_bgk.dat};
\legend{{spectral, same stencils}, {$\mathrm{RR}_{\omega,1}$, $N=512$}, {$\mathrm{RR}_{\omega,1}$, $N=256$}, {RR, $N=512$}, {RR, $N=256$}, {BGK, $N=512$}, {BGK, $N=256$}}
\addplot[cRR, only marks, mark=x, mark size=3pt, solid, forget plot] coordinates {(0.5499,18.2171)};
\addplot[cCM, only marks, mark=x, mark size=3pt, solid, forget plot] coordinates {(0.7486,39.2784)};
\addplot[cBGK, only marks, mark=x, mark size=3pt, solid, forget plot] coordinates {(0.1500,5.9596)};
\addplot[cBGK!50, only marks, mark=x, mark size=3pt, solid, forget plot] coordinates {(0.2495,7.2752)};
\nextgroupplot[title={(b) $\max|\nabla\cdot\mathbf b|\,/\max|\mathbf b|$}, ymode=log, ymin=1e-5, ymax=20]
\addplot[cRR, thick] table[x=t, y=divb] {data/gpu_re3000_m512_cm_wb1.dat};
\addplot[cRRL, thick] table[x=t, y=divb] {data/gpu_re3000_m256_cm_wb1.dat};
\addplot[cRR, thick, densely dashed] table[x=t, y=divb] {data/gpu_re3000_m512_cm_wbomega.dat};
\addplot[cCM, thick, dashdotted] table[x=t, y=divb] {data/gpu_re3000_m256_cm_wbomega.dat};
\addplot[cBGK, thick, densely dotted] table[x=t, y=divb] {data/gpu_re3000_m512_bgk.dat};
\addplot[cBGK!50, thick, densely dotted] table[x=t, y=divb] {data/gpu_re3000_m256_bgk.dat};
\addplot[cRR, only marks, mark=x, mark size=3pt, solid, forget plot] coordinates {(0.5499,0.1101)};
\addplot[cCM, only marks, mark=x, mark size=3pt, solid, forget plot] coordinates {(0.7486,0.3000)};
\addplot[cBGK, only marks, mark=x, mark size=3pt, solid, forget plot] coordinates {(0.1500,0.0382)};
\addplot[cBGK!50, only marks, mark=x, mark size=3pt, solid, forget plot] coordinates {(0.2495,0.4727)};
\end{groupplot}
\node[anchor=south] at ($(group c1r1.north)!0.5!(group c2r1.north)+(0,0.85cm)$) {\pgfplotslegendfromname{leg:highre}};
\end{tikzpicture}
\caption{Three-dimensional vortex at $\mathrm{Re}=3000$: (a) $J_{\max}$ and (b) $\max|\nabla\cdot\bb|/\max|\bb|$ (physical units) for $\mathrm{RR}_{\omega,1}$ and RR on $N=256$ and $N=512$, and BGK on both grids. Grey bands in (a): pseudo-spectral solution ($N_s=512$) evaluated with the stencil of each grid on its nodes, $N=512$ the darker. Crosses mark the last sample before the blow-up; the divergence times of Table~\ref{tab:div} are the first non-finite samples, one or two samples later.}
\label{fig:highre}
\end{figure}

The failure starts in the compressive modes of the fluid (Fig.~\ref{fig:mech}a). On $N=128$ at $\mathrm{Re}=3000$, single-rate RR and $\mathrm{RR}_{\omega,1}$ agree to $0.6\%$ in $J_{\max}$, vorticity and $\max|\nabla\cdot\bb|$ until $t=1.38$. Then $\max|\nabla\cdot\bu|$ of single-rate RR doubles within $\Delta t=0.07$ while the others stay within $0.5\%$; $\nabla\cdot\bb$ follows at $t\approx1.45$ and the current only at $1.64$, before the divergence at $1.78$. The finer grids show the same order with shorter delays, in runs sampled every $0.005$. Since $\omega_b=1$ damps the sound waves of the start, the ratio of $\max|\nabla\cdot\bu|$ between the two rates fluctuates there from the beginning, up to $2.7$ on $N=256$ and $1.6$ on $N=512$; it stays above $3$ from $t=0.69$ and $0.51$, when $J_{\max}$ still agrees to $0.03\%$ and $0.02\%$. $\max|\nabla\cdot\bb|$ departs by more than $5\%$ from $t=0.74$ on $N=256$ and, together with the divergence of the velocity, from $0.51$ on $N=512$; the current departs by more than $1\%$ from $0.76$ and $0.56$, and the runs diverge at $0.82$ and $0.63$. These modes are carried by the trace of the second-order non-equilibrium, which with $\omega_b=\omega\approx2$ is reflected rather than damped at each collision and with $\omega_b=1$ is removed at every step. A scan of the bulk rate (Fig.~\ref{fig:mech}b) shows how much damping is needed: with $\nu_b\approx20\nu$ and $42\nu$ ($\omega_b=1.9$ and $1.8$) the runs diverge as the single rate does, with $125\nu$ ($1.5$) the run survives a local burst, and only $\omega_b=1$, $\nu_b\approx375\nu$, avoids the departure. The $\nabla\cdot\bb$ that grows on the way is the central-difference divergence of the nodal field, a truncation error that adds a force $\bb(\nabla\cdot\bb)$ to a momentum equation in conservation form~\cite{brackbill1980,toth2000}; the divergence that the scheme controls, the trace of~\eqref{eq:gradb}, stays below $10^{-13}$~\cite{dellar2002}.

\begin{figure}[!htbp]
\centering
\begin{tikzpicture}
\begin{groupplot}[group style={group size=2 by 2, horizontal sep=1.65cm, vertical sep=3.5cm}, width=0.47\linewidth, height=5.6cm, grid=major, grid style={gray!20}, legend style={draw=none, fill=white, fill opacity=0.85, text opacity=1, font=\footnotesize}, legend cell align=left]
\nextgroupplot[title={(a) single-rate RR / $\mathrm{RR}_{\omega,1}$}, xlabel={$t$}, xmin=1.0, xmax=1.75, ymode=log, ymin=0.8, ymax=30, ytick={1,2,5,10,20}, yticklabels={$1$,$2$,$5$,$10$,$20$}, legend pos=north west]
\addplot[cCM, thick] table[x=t, y=divU] {data/mech_ratio.dat};
\addplot[cRR, thick, densely dashed] table[x=t, y=divB] {data/mech_ratio.dat};
\addplot[cBGK, thick, densely dotted] table[x=t, y=J] {data/mech_ratio.dat};
\addplot[cRM, thick, dashdotted] table[x=t, y=W] {data/mech_ratio.dat};
\legend{$\max|\nabla\cdot\mathbf u|$, $\max|\nabla\cdot\mathbf b|$, $J_{\max}$, $\max|\boldsymbol\omega|$}
\nextgroupplot[title={(b) $\max|\nabla\cdot\mathbf u|$, bulk rate $\omega_b$}, xlabel={$t$}, xmin=0, xmax=2, ymode=log, ymin=0.02, ymax=500, legend style={at={(0.5,-0.3)}, anchor=north, font=\scriptsize}, legend columns=3]
\addplot[cRR, thick, densely dashed] table[x=t, y=divU] {data/mech_n128_wbomega.dat};
\addplot[cPR, thick] table[x=t, y=divU] {data/mech_n128_wb1.9.dat};
\addplot[cRM, thick] table[x=t, y=divU] {data/mech_n128_wb1.8.dat};
\addplot[cRRL, thick] table[x=t, y=divU] {data/mech_n128_wb1.5.dat};
\addplot[cRR, thick] table[x=t, y=divU] {data/mech_n128_wb1.0.dat};
\addplot[black, thick, densely dotted] table[x=t, y=divU] {data/mech_n128_halfv0.dat};
\legend{{$\omega_b=\omega$, $0.67\nu$}, {$\omega_b=1.9$, $20\nu$}, {$\omega_b=1.8$, $42\nu$}, {$\omega_b=1.5$, $125\nu$}, {$\omega_b=1$, $375\nu$}, {$\omega_b=\omega$, $v_0/2$}}
\addplot[cRR, only marks, mark=x, mark size=3pt, solid, forget plot] coordinates {(1.7681,103.2931)};
\addplot[cPR, only marks, mark=x, mark size=3pt, solid, forget plot] coordinates {(1.7886,132.0578)};
\addplot[cRM, only marks, mark=x, mark size=3pt, solid, forget plot] coordinates {(1.8192,245.8077)};
\nextgroupplot[title={(c) linear growth per step, $N=128$}, xlabel={$|\mathbf B_0|$ (lattice units)}, ymode=log, xmin=0, xmax=0.046, ymin=1e-5, ymax=0.05, xtick={0,0.01,0.02,0.03,0.04}, scaled x ticks=false, xticklabel style={/pgf/number format/fixed, /pgf/number format/precision=3}, legend style={at={(0.5,-0.3)}, anchor=north, font=\scriptsize}, legend columns=3, unbounded coords=jump]
\fill[gray!15] (axis cs:0.0136,1e-5) rectangle (axis cs:0.0254,0.05);
\addplot[cBGK, thick, densely dotted] table[x=B0, y=g_bgk] {data/lsa_growth_classes.dat};
\addplot[cRR, thick, densely dashed] table[x=B0, y=gc_omega] {data/lsa_growth_classes.dat};
\addplot[cPR, thick] table[x=B0, y=gc_1.9] {data/lsa_growth_classes.dat};
\addplot[cRM, thick] table[x=B0, y=gc_1.8] {data/lsa_growth_classes.dat};
\addplot[cRRL, thick] table[x=B0, y=gc_1.5] {data/lsa_growth_classes.dat};
\addplot[cRR, thick] table[x=B0, y=gc_1.0] {data/lsa_growth_classes.dat};
\addplot[black, thin, dashdotted] table[x=B0, y=go_1.0] {data/lsa_growth_classes.dat};
\legend{{BGK}, {$\omega_b=\omega$}, {$\omega_b=1.9$}, {$\omega_b=1.8$}, {$\omega_b=1.5$}, {$\omega_b=1$}, {slow mode}}
\nextgroupplot[title={(d) instability threshold, $\mathrm{Re}=3000$}, xlabel={$N$}, ylabel={$|\mathbf B_0|$ (lattice units)}, xmode=log, log basis x=2, xmin=48, xmax=680, ymin=0, ymax=0.08, xtick={64,128,256,512}, xticklabels={64,128,256,512}, scaled y ticks=false, yticklabel style={/pgf/number format/fixed, /pgf/number format/precision=2}, legend style={at={(0.5,-0.3)}, anchor=north, font=\scriptsize}, legend columns=3]
\addplot[cRR, thick, densely dashed, mark=square*] table[x=N, y=B_omega, restrict expr to domain={\thisrow{Re}}{2999:3001}] {data/lsa_thresholds_flow.dat};
\addplot[cPR, thick, mark=triangle*] table[x=N, y=B_1.9, restrict expr to domain={\thisrow{Re}}{2999:3001}] {data/lsa_thresholds_flow.dat};
\addplot[cRRL, thick, mark=diamond*] table[x=N, y=B_1.5, restrict expr to domain={\thisrow{Re}}{2999:3001}] {data/lsa_thresholds_flow.dat};
\addplot[cRR, thick, mark=*] table[x=N, y=B_1.0, restrict expr to domain={\thisrow{Re}}{2999:3001}] {data/lsa_thresholds_flow.dat};
\addplot[cRR, thick, densely dotted, mark=o, mark options={solid}] table[x=N, y=B] {data/lsa_compressive_wb1.dat};
\addplot[black, dashed, domain=48:680, samples=2] {0.0254};
\addplot[black, dotted, domain=48:680, samples=2] {0.0136};
\legend{{$\omega_b=\omega$}, {$\omega_b=1.9$}, {$\omega_b=1.5$}, {$\omega_b=1$}, {$\omega_b=1$, compressive}, {peak initial $|\mathbf b|$}, {rms initial $|\mathbf b|$}}
\end{groupplot}
\end{tikzpicture}
\caption{Three-dimensional vortex at $\mathrm{Re}=3000$ on $N=128$. (a) Ratio of single-rate RR to $\mathrm{RR}_{\omega,1}$ of $\max|\nabla\cdot\bu|$, $\max|\nabla\cdot\bb|$ (both central differences), $J_{\max}$ and the peak vorticity: the compressive part of the velocity departs first. (b) $\max|\nabla\cdot\bu|$ (physical units) for bulk rates between $\omega_b=\omega\approx1.996$ and $\omega_b=1$, with the bulk viscosity $\nu_b$ in units of the shear viscosity, and for the single rate at half the lattice velocity (dotted); crosses mark the last finite sample before divergence. (c) Growth per time step of linear perturbations of a uniform state with field $\mathbf B_0$ and no mean flow, at the viscosity of these runs (\ref{app:lsa}): the compressive mode for the bulk rates of (b), the most unstable mode of BGK, and the slow non-compressive mode (Table~\ref{tab:lsa}), which exists for every bulk rate and is drawn for $\omega_b=1$; the band spans the initial field of the runs, from its rms to its peak value. (d) Field above which the uniform state with a mean flow $|\mathbf U|=0.02$ is unstable, against the grid at $\mathrm{Re}=3000$; open circles: threshold of the compressive mode alone for $\omega_b=1$; dashed and dotted lines: peak and rms initial field.}
\label{fig:mech}
\end{figure}

A linear analysis of the coupled scheme about a uniform state explains why (\ref{app:lsa}, Table~\ref{tab:lsa}, Fig.~\ref{fig:mech}c,d). The 27 fluid and 21 magnetic populations are linearised about a state with uniform field $\mathbf B_0$ and velocity $\mathbf U$, and the eigenvalues of collision followed by streaming give the growth of every Fourier mode. Without a field RR is stable for every bulk rate, even with a mean flow at $\mathrm{Ma}\approx0.09$. A uniform field destabilises a grid-scale compressive mode propagating across it, at $k\Delta x\approx2.6$ along a lattice axis. At the viscosity of the $N=128$ runs its threshold is $|\mathbf B_0|=0.0037$ in lattice units for the single rate, with or without a mean flow of the vortex's size, and the bulk viscosity raises it to $0.038$ for $\omega_b=1$: the initial field of the runs, $0.014$ rms and $0.025$ at its peak, exceeds the first almost everywhere and the second nowhere. The same holds on every grid, where the compressive threshold of $\omega_b=1$ stays above the peak initial field ($0.037$--$0.072$ on $N=64$--$512$) and that of the single rate below it ($0.0018$--$0.014$, in proportion to the lattice viscosity). These are thresholds of the unbounded lattice: near them the unstable band is narrower than the spacing of the wavenumbers of the periodic box, whose first wavevector enters it at $|\mathbf B_0|\approx0.012$ on $N=128$ (\ref{app:lsa}). On $N=128$, $256$ and $512$ the single rate fails earlier where the growth at the peak field is larger, $38$, $62$ and $73$ e-folds per unit time, although $N=64$, with $21$, fails slightly before $N=128$. Linear instability does not decide alone, however: at $\mathrm{Re}=1000$ the single rate is also unstable at the peak field and runs to $t=3$, and with $\omega_b=1$ the scheme fails on $N=64$ at $\mathrm{Re}=3000$ and on $N=256$ and $512$ at $\mathrm{Re}=10\,000$ although its compressive thresholds there, $0.037$--$0.039$, lie above the peak initial field. The linear analysis therefore accounts for four things: the mode that fails, a grid-scale compressive one; the order of the collisions, BGK first, then the single rate, then $\omega_b=1$; the size of the thresholds in relation to the field of the runs; and the action of the bulk viscosity, which damps the very mode that the Maxwell coupling destabilises. It concerns uniform states, and gives neither the divergence times nor the dependence on the lattice velocity described below.

\begin{figure}[!htbp]
\centering
\begin{tikzpicture}
\begin{groupplot}[group style={group size=2 by 1, horizontal sep=2.0cm}, width=0.47\linewidth, height=5.6cm, grid=major, grid style={gray!20}, legend style={draw=none, fill=white, fill opacity=0.85, text opacity=1, font=\scriptsize}, legend cell align=left]
\nextgroupplot[title={(a) divergence time}, xlabel={lattice Mach number $\mathrm{Ma}$}, ylabel={$t$}, xmode=log, xmin=0.015, xmax=0.12, xtick={0.017,0.024,0.034,0.051,0.068,0.102}, xticklabels={$0.017$,$0.024$,$0.034$,$0.051$,$0.068$,$0.10$}, xticklabel style={font=\scriptsize}, ymin=1.15, ymax=3.42, ytick={1.2,1.6,2,2.4,2.8}, legend style={at={(0.5,-0.32)}, anchor=north}, legend columns=2]
\addplot[cN64, densely dashed, thick, mark=o, mark options={solid}] coordinates {(0.02210,2.3225) (0.02380,1.9431) (0.02720,1.7813) (0.03060,1.7097) (0.03400,1.6557)};
\addplot[cN64, only marks, mark=triangle, mark size=3pt, mark options={solid}, forget plot] coordinates {(0.01700,3.0800) (0.02040,3.0800)};
\addplot[cN64, thick, mark=*, mark options={solid}] coordinates {(0.02550,2.0543) (0.02720,1.8849) (0.03060,1.7722) (0.03400,1.7068) (0.05100,1.5535) (0.06800,1.4281) (0.10200,1.2571)};
\addplot[cN64, only marks, mark=triangle*, mark size=3pt, mark options={solid}, forget plot] coordinates {(0.02040,3.1600) (0.02380,3.1600)};
\addplot[cN128, densely dashed, thick, mark=square, mark options={solid}] coordinates {(0.03400,1.7783)};
\addplot[cN128, only marks, mark=triangle, mark size=3pt, mark options={solid}, forget plot] coordinates {(0.01700,3.2700) (0.02550,3.2700)};
\addplot[cN128, thick, mark=square*, mark options={solid}] coordinates {(0.05100,1.4717) (0.06800,1.3389)};
\addplot[cN128, only marks, mark=triangle*, mark size=3pt, mark options={solid}, forget plot] coordinates {(0.03400,3.3500)};
\draw[gray, densely dotted] (axis cs:0.015,3.02) -- (axis cs:0.12,3.02);
\node[font=\scriptsize, gray, anchor=north east] at (axis cs:0.118,3.0) {above: runs to $t=3$};
\legend{{$N=64$, single rate}, {$N=64$, $\omega_b=1$}, {$N=128$, single rate}, {$N=128$, $\omega_b=1$}}
\nextgroupplot[title={(b) $\max|\nabla\cdot\mathbf u|/\mathrm{Ma}$, $N=128$}, xlabel={$t$}, xmin=0.8, xmax=3, ymode=log, ymin=10, ymax=3000, legend style={at={(0.5,-0.32)}, anchor=north}, legend columns=2]
\addplot[cRR, densely dashed, thick] table[x=t, y expr=\thisrow{divU}/0.017] {data/vscan_n128_f0.5_omega.dat};
\addplot[cRM, densely dashed, thick] table[x=t, y expr=\thisrow{divU}/0.0255] {data/vscan_n128_f0.75_omega.dat};
\addplot[cCM, densely dashed, thick] table[x=t, y expr=\thisrow{divU}/0.034] {data/vscan_n128_f1.0_omega.dat};
\addplot[cCM, only marks, mark=x, mark size=3pt, forget plot] coordinates {(1.7681,2900.0000)};
\addplot[cCM, thick] table[x=t, y expr=\thisrow{divU}/0.034] {data/vscan_n128_f1.0_wb1.dat};
\addplot[black, thick] table[x=t, y expr=\thisrow{divU}/0.051] {data/vscan_n128_f1.5_wb1.dat};
\addplot[black, only marks, mark=x, mark size=3pt, forget plot] coordinates {(1.4615,2494.9698)};
\addplot[black, densely dotted, thick] table[x=t, y expr=\thisrow{divU}/0.068] {data/vscan_n128_f2.0_wb1.dat};
\addplot[black, only marks, mark=x, mark size=3pt, forget plot] coordinates {(1.3286,2900.0000)};
\legend{{single, $\mathrm{Ma}=0.017$}, {single, $\mathrm{Ma}=0.026$}, {single, $\mathrm{Ma}=0.034$}, {$\omega_b=1$, $\mathrm{Ma}=0.034$}, {$\omega_b=1$, $\mathrm{Ma}=0.051$}, {$\omega_b=1$, $\mathrm{Ma}=0.068$}}
\end{groupplot}
\end{tikzpicture}
\caption{Lattice-velocity scan of the three-dimensional vortex at $\mathrm{Re}=3000$, at fixed $N$. (a) Divergence time against the lattice Mach number for the single rate (open markers, dashed) and $\omega_b=1$ (filled markers, solid) on $N=64$ and $128$; triangles above the dotted line are runs that reach $t=3$, one row per grid and rate ($N=128$ above). (b) $\max|\nabla\cdot\bu|$ divided by $\mathrm{Ma}$ on $N=128$; crosses mark the last sample before divergence.}
\label{fig:vscan}
\end{figure}

\begin{figure}[!htb]
\centering
\input{figs/fig_iso3dgpu}
\caption{Isosurfaces of the current density $|\mathbf j|$ at $40\%$ of the maximum of the displayed field: RR at $\mathrm{Re}=1000$ on $N=256$ (a, b) and $\mathrm{RR}_{\omega,1}$ at $\mathrm{Re}=3000$ on $N=512$ (c, d), at $t\approx1$ (a, c) and $t\approx2$ (b, d). $\mathbf j$ is computed from $\bb$ by spectral differentiation over $|k_\alpha|\le\min(N/3,N/(2s)-1)$ on every $s$-th node, i.e.\ $|k_\alpha|\le85$ for $N=256$ ($s=1$) and $|k_\alpha|\le127$ for $N=512$ ($s=2$), and smoothed with the Gaussian filter of Fig.~\ref{fig:iso3d} (width $2\pi/128$).}
\label{fig:iso3dgpu}
\end{figure}

\begin{table}[!htbp]
\centering\footnotesize
\caption{Controls: divergence times, defined as in Table~\ref{tab:div}, of the six collisions on the three smallest cases (samples every $0.01$). --: runs to the end ($t=2$; $t=3$ for $\mathrm{RR}_{\omega,1}$ on $N=128$).}
\label{tab:ctrl}
\begin{tabular}{@{}lccc@{}}
\toprule
 & $\mathrm{Re}=1000$ & \multicolumn{2}{c}{$\mathrm{Re}=3000$}\\
\cmidrule(lr){2-2}\cmidrule(l){3-4}
$N$ & 64 & 64 & 128\\
\midrule
BGK & 0.79 & 0.75 & 0.51\\
$\mathrm{BGK}_{\omega,1}$ & 1.18 & 1.04 & 0.58\\
PR & -- & 1.65 & 1.76\\
$\mathrm{PR}_{\omega,1}$ & -- & 1.70 & --\\
RR & -- & 1.66 & 1.78\\
$\mathrm{RR}_{\omega,1}$ & -- & 1.71 & --\\
\bottomrule
\end{tabular}
\end{table}

Two controls separate the ingredients (Table~\ref{tab:ctrl}). BGK goes unstable first, from $|\mathbf B_0|\approx0.0022$, through modes that are mostly non-hydrodynamic. With $\omega_b=1$ its threshold rises only to $0.0028$, where a mode nearly along the field, which the trace does not carry, becomes unstable, and the runs still diverge early: at $\mathrm{Re}=3000$ at $t\approx0.58$ against $0.51$ on $N=128$ and at $1.04$ against $0.75$ on $N=64$, and at $\mathrm{Re}=1000$ at $1.18$ against $0.79$ on $N=64$. Projective regularisation has the thresholds and unstable modes of RR, and behaves like it: on $N=128$ single-rate PR diverges at $t\approx1.76$, and $\mathrm{PR}_{\omega,1}$ runs to $t=2$ within $0.5\%$ of $\mathrm{RR}_{\omega,1}$ in $J_{\max}$; on $N=64$, where the regularised schemes fail with either bulk rate at this lattice velocity, PR and RR diverge within one sample of each other. Robustness therefore needs both the regularisation and the unit bulk rate, whereas the frame of the filter does not matter. In all these runs, and in the linearised operator, the magnetic populations keep their BGK collision; its share in the thresholds has not been separated, so a contribution of the magnetic collision to these stability limits cannot be excluded.

The lattice velocity decides as well (Fig.~\ref{fig:vscan}). At fixed $N$ and $\mathrm{Re}$, changing $v_0$ changes only the lattice Mach number $\mathrm{Ma}$, since the grid Reynolds number $\mathrm{Re}/N$ is fixed, and the uniform-state analysis is invariant under this change to within $2\%$ in the growth per unit time; only the width of its unstable band changes, in proportion to $v_0$ (\ref{app:lsa}). The runs are not: on $N=64$ and $128$ each scheme fails only above a critical lattice Mach number. For the single rate it lies between $0.020$ and $0.022$ on $N=64$ and between $0.026$ and $0.034$ on $N=128$; with $\omega_b=1$ it rises to between $0.024$ and $0.026$, and between $0.034$ and $0.051$. These brackets are those of this flow, Reynolds number and grid, not properties of the scheme: at the same $\mathrm{Ma}=0.034$ the single rate also fails on $N=256$ and $512$, earlier than on $N=128$ (Table~\ref{tab:div}), and on $N=256$ no threshold was found down to $\mathrm{Ma}=0.017$: there the single rate still diverges, at $t\approx1.02$, while $\omega_b=1$ runs to the end of that run, $t=2$. Above the critical value the failure comes earlier as $\mathrm{Ma}$ grows, on $N=64$ from $t\approx2.32$ at $\mathrm{Ma}=0.022$ to $1.66$ at $0.034$ for the single rate and from $2.05$ at $0.026$ to $1.26$ at $0.10$ with $\omega_b=1$; below it the runs pass the current peak and reach $t=3$. Until the failing runs depart, $\max|\nabla\cdot\bu|$ grows with $\mathrm{Ma}$, less than in proportion before $t\approx1.2$, and hardly depends on the bulk rate: at $\mathrm{Ma}=0.034$ the two rates give the same history to within $20\%$ up to $t\approx1.35$. The failing runs then leave the band of the others and diverge a few tenths of a time unit later, the earlier the higher $\mathrm{Ma}$: they leave it at $t\approx1.1$ for $\omega_b=1$ at $\mathrm{Ma}=0.068$, $1.3$ at $0.051$, and $1.4$ for the single rate at $0.034$ (Fig.~\ref{fig:vscan}b). The bulk rate acts on this departure, not on a seed left by the uniform-density start: with $\omega_b=1$ only before $t=1$ the run at $\mathrm{Ma}=0.034$ still departs at $t\approx1.45$ and fails, whereas switching from the single rate to $\omega_b=1$ at $t=1.3$, just before the departure, arrests it: $\max|\nabla\cdot\bu|$ peaks at $t\approx1.75$ at twice the level of $\mathrm{RR}_{\omega,1}$, is back to it by $t\approx2.1$, and the run reaches $t=2.5$. The instability thus depends on the lattice Mach number through the amplitude of the compressive modes or their time scale relative to the flow, which an analysis of uniform states does not contain: that analysis ranks the schemes but cannot place the threshold, which remains unexplained. At half the lattice velocity it predicts an unstable single rate, with $37$ and $62$ e-folds per unit time at the peak field on $N=128$ and $256$, and a stable $\omega_b=1$: the runs bear this out on $N=256$ and not on $N=128$, where the single rate survives. At $\mathrm{Re}=10\,000$ a lower lattice velocity does not help: at $\mathrm{Ma}=0.017$ the single rate diverges at $t\approx0.76$ on $N=256$ and $0.40$ on $N=512$, against $0.65$ and $0.35$ at $\mathrm{Ma}=0.034$, and $\omega_b=1$ at $1.38$ and $1.14$, against $1.10$ and $0.90$: later, but still before the current peak, although the compressive mode of $\omega_b=1$ is linearly stable on both grids at both lattice velocities.

The brackets move with the configuration as well. A uniform guide field $3\hat{\mathbf z}$ added to the initial field (the guide-field configuration of \ref{app:lorentz}) raises it in lattice units, at $\mathrm{Ma}=0.034$, from $0.014$ to $0.025$ rms and from $0.025$ to $0.039$ at its peak. At $\mathrm{Re}=3000$ on $N=128$ single-rate RR then diverges at $t\approx1.11$, $1.20$ and $1.38$ at $\mathrm{Ma}=0.034$, $0.022$ and $0.017$, and $\mathrm{RR}_{\omega,1}$ at $1.49$ and $1.93$ at the first two; at $\mathrm{Ma}=0.017$ it runs to $t=3$ and, continued to $t=4$, passes a second maximum of $J_{\max}$, $160$ at $t\approx3.15$ against $134$ at $1.90$, and comes back down. With the guide field the critical lattice Mach number of $\omega_b=1$ on this grid thus lies between $0.017$ and $0.022$, against $0.034$--$0.051$ without it, and none was found for the single rate down to $0.017$. On $N=64$ none was found for either rate: the six runs diverge between $t\approx1.56$ and $2.12$, the unit bulk rate delaying the failure by at most $0.09$, and by one sample at $\mathrm{Ma}=0.022$. On $N=128$ the single-rate failure is the one described above: the two rates agree to $0.5\%$ in $J_{\max}$ until $\max|\nabla\cdot\bu|$ of the single rate departs, at $t\approx0.98$, $1.01$ and $1.17$ at the three lattice velocities, before $\max|\nabla\cdot\bb|$ and the current. The uniform-state analysis is consistent with the earlier failure of the single rate, whose growth at the peak field rises from about $38$ to $60$--$62$ e-folds per unit time on $N=128$, but again places no threshold. That growth is the same at the three lattice velocities, while the divergence time moves from $1.11$ to $1.38$. The compressive threshold of $\omega_b=1$, $0.037$--$0.038$, lies just below the peak field at $\mathrm{Ma}=0.034$, where the growth is under two e-folds per unit time and the run fails at $t\approx1.49$, and above it at $\mathrm{Ma}=0.022$ and $0.017$ (peak field $0.025$ and $0.019$), where one run fails and the other does not. Nor does the peak field in lattice units decide: at $\mathrm{Ma}=0.022$ it equals that of the standard field at $\mathrm{Ma}=0.034$, with which $\mathrm{RR}_{\omega,1}$ runs to $t=3$ on $N=128$. No reference solution was computed for this configuration at $\mathrm{Re}=3000$, and finer grids were not run: these runs test robustness on $N\le128$, not accuracy.

Figure~\ref{fig:iso3dgpu} shows the sheets of single-rate RR at $\mathrm{Re}=1000$ and of $\mathrm{RR}_{\omega,1}$ at $\mathrm{Re}=3000$: alike at $t\approx1$, and broken into many smaller sheets by $t\approx2$, more so at $\mathrm{Re}=3000$.

\section{Conclusions}\label{sec:conclusions}

The answers to the three questions posed in the Introduction are as follows.
\begin{itemize}
\item Complete recursive regularisation is a central-Hermite-moment collision on D2Q9 and D3Q27, for any common second-order relaxation map and any reference equilibrium with the prescribed density and momentum. The closed form $R_i(A)=\tfrac{w_i}{2}A_{\alpha\beta}\partial^2F_i/\partial u_\alpha\partial u_\beta$ gives the regularised populations directly from the one-dimensional factors of the product-form function. The central moments of the Maxwell stress are the corresponding recursion coefficients, up to the sign $(-1)^{|n|}$. A central-moment implementation must therefore retain these contributions through the highest represented order or, if it omits them, use the corresponding comoving Maxwell equilibrium. Two-dimensional raw- and central-moment schemes with the trace relaxed to equilibrium correspond to projective and recursive regularisation with a unit bulk rate, respectively. On D3Q27, an independent CHM implementation reproduces RR to round-off, whereas the tested monomial collision differs at fourth order and produces a larger kinetic-energy error in the three-dimensional vortex considered here.
\item The Chapman--Enskog analysis shows that the hydrodynamic recursion misses most of the third-order non-equilibrium contribution generated by the Maxwell stress in the tested magnetic configurations. A local correction restores this contribution, but its effect on the macroscopic solutions considered here is smaller than the discretisation error, consistent with its Burnett-order character. Dellar's Maxwell equilibrium also introduces a frame-dependent contribution to the viscous stress. For the standard field this is smaller than the cubic defect of the lattice, although it becomes more significant as the magnetic field and local velocity increase.
\item The D3Q27 scheme converges at approximately second order at $\mathrm{Re}=100$ and follows the pseudo-spectral reference through the peak-current stage at $\mathrm{Re}=1000$ and, with a unit bulk rate, at $\mathrm{Re}=3000$. At $\mathrm{Re}=3000$, single-rate RR develops a compressive instability whose observed threshold depends on the grid and initial magnetic configuration. For the standard initial field, introducing a unit bulk rate allows the simulations at $\mathrm{Ma}=0.034$ to reach the end of the run for $N\ge128$, without measurable loss of accuracy at $\mathrm{Re}=1000$. The guide-field configuration is less stable: on $N=128$, the critical Mach number for the unit bulk rate lies between $0.017$ and $0.022$. The combined strategy does not prevent failure at $\mathrm{Re}=10,000$, even at half the lattice velocity. Linear stability analysis identifies the unstable mode and explains the role of bulk dissipation, but it does not predict the nonlinear Mach-number threshold or failure time.
\end{itemize}

The central message is that regularisation, moment representation and bulk dissipation are distinct ingredients of a robust MHD lattice Boltzmann scheme. The regularisation removes the non-hydrodynamic content of the populations; the moment basis decides what is held at equilibrium beyond second order; the bulk rate controls the compressive modes that the Maxwell coupling destabilises. None of them replaces another. Natural next steps are wall-bounded MHD flows, entropic and cumulant collisions with a separate bulk rate, the role of the magnetic collision in the stability limits, and a theory of the critical lattice Mach number.

\appendix
\section{First-order Chapman--Enskog non-equilibrium}\label{app:ce}
\setcounter{table}{0}\setcounter{figure}{0}
\renewcommand{\theHfigure}{appA.\arabic{figure}}
\renewcommand{\theHtable}{appA.\arabic{table}}

This appendix derives~\eqref{eq:a1} and~\eqref{eq:f1}. With $\tau=1/\omega$, the first-order Chapman--Enskog non-equilibrium is $f^{(1)}_i=-\tau(\partial^{(1)}_t+c_{i\gamma}\partial_\gamma)f^\eq_i$. For any population vector $g$ let $a_m[g]=\sum_iH_{\boldsymbol m}(\bc_i)g_i$, so that $a_{1,n}=a_n[f^{(1)}]$. On D1Q3, $c\He_p(c)=\He_{p+1}(c)+p\,\cs\He_{p-1}(c)$ for $p=0,1,2$, with $\He_3\equiv0$ because $c^3=c$. Hence
\begin{equation}\label{eq:shift}
\sum_iH_{n}(\bc_i)\,c_{i\gamma}\,g_i=[n_\gamma\le1]\,a_{n+\ee_\gamma}[g]+n_\gamma\cs\,a_{n-\ee_\gamma}[g],
\end{equation}
where $[\cdot]$ is one when the condition holds and zero otherwise. The equilibrium~\eqref{eq:feq} has $a_{\boldsymbol m}[f^\eq]=\rho\bu^{\boldsymbol m}+M_{\boldsymbol m}$, with $M_{\boldsymbol m}=M_{\alpha\beta}$ for $\boldsymbol m=\ee_\alpha+\ee_\beta$ and $M_{\boldsymbol m}=0$ for $|\boldsymbol m|\ne2$. Therefore
\begin{multline}\label{eq:a1n}
a_{1,n}=-\tau\Big[\partial_t(\rho\bu^{n})+\sum_\gamma\partial_\gamma(\rho\bu^{n}u_\gamma)+\cs\sum_\gamma n_\gamma\partial_\gamma(\rho\bu^{n-\ee_\gamma})\Big]\\
-\tau\Big[\partial_tM_{n}+\cs\sum_\gamma n_\gamma\partial_\gamma M_{n-\ee_\gamma}\Big]+\tau\sum_{\gamma:\,n_\gamma=2}\partial_\gamma(\rho\bu^{n}u_\gamma),
\end{multline}
where the last term removes from the first bracket the streaming contributions with $n_\gamma=2$, which~\eqref{eq:shift} does not contain. In the first bracket, the ideal continuity and momentum equations give $\partial_t(\rho\bu^{n})+\nabla\cdot(\rho\bu^{n}\bu)=\rho(\partial_t+\bu\cdot\nabla)\bu^{n}=\sum_\gamma n_\gamma\bu^{n-\ee_\gamma}\big[-\partial_\gamma(\rho\cs)-\partial_\beta M_{\gamma\beta}\big]$. The density gradient cancels against the part $\cs\sum_\gamma n_\gamma\bu^{n-\ee_\gamma}\partial_\gamma\rho$ of the third term, and the bracket reduces to $\rho\cs\sum_\gamma n_\gamma\partial_\gamma\bu^{n-\ee_\gamma}-\sum_\gamma n_\gamma\bu^{n-\ee_\gamma}\partial_\beta M_{\gamma\beta}$. In the index notation of~\eqref{eq:a1}, $\sum_\gamma n_\gamma\partial_\gamma\bu^{n-\ee_\gamma}=\sum_{l<m}(\partial_{\alpha_l}u_{\alpha_m}+\partial_{\alpha_m}u_{\alpha_l})\prod_{r\ne l,m}u_{\alpha_r}$ and $\sum_\gamma n_\gamma\bu^{n-\ee_\gamma}\partial_\beta M_{\gamma\beta}=\sum_l\partial_\beta M_{\alpha_l\beta}\prod_{r\ne l}u_{\alpha_r}$, which give the first two terms of~\eqref{eq:a1}. The second bracket of~\eqref{eq:a1n} is $\partial_tM_{n}$ for $|n|=2$ and $\cs T_{n}$ for $|n|=3$, since $\sum_\gamma n_\gamma\partial_\gamma M_{n-\ee_\gamma}$ is the multi-index form of $T_{\alpha_1\alpha_2\alpha_3}$; it vanishes for $|n|\ge4$. This gives~\eqref{eq:a1}. The last term of~\eqref{eq:a1n} exists only for components with a repeated index; for $n=2\ee_\alpha$ it is the cubic defect $\tau\partial_\alpha(\rho u_\alpha^3)$ of the viscous stress~\cite{dellar2014}.

For the closed form, the pair sum of $a^{\rm h}$ in~\eqref{eq:a1} is $\RecA_{n}[a^{\rm h}]$, whose Hermite series is $R_i(a^{\rm h})$ by~\eqref{eq:closed}. The Lorentz term is $\tau\partial_\beta M_{\gamma\beta}$ times $\partial\bu^{n}/\partial u_\gamma$ for $|n|\ge2$, and by~\eqref{eq:generating}
\begin{equation}
w_i\sum_{n}\frac{\partial\bu^{n}/\partial u_\gamma}{n!\,c_s^{2|n|}}H_{n}(\bc_i)=w_i\frac{\partial F_i}{\partial u_\gamma},
\end{equation}
whose first-order part is $w_ic_{i\gamma}/\cs$; removing it gives the second term of~\eqref{eq:f1}. At second and third order the Maxwell terms have the coefficients $-\tau\partial_tM$ and $-\tau\cs T$, and $T_{\alpha\beta\gamma}\mathcal H^{(3)}_{i,\alpha\beta\gamma}=3\,\mathcal H^{(3)}_{i,\alpha\beta\gamma}\partial_\gamma M_{\alpha\beta}$ gives the last term.

\section{Pseudo-spectral reference solutions}\label{app:spectral}
\setcounter{table}{0}\setcounter{figure}{0}
\renewcommand{\theHfigure}{appB.\arabic{figure}}
\renewcommand{\theHtable}{appB.\arabic{table}}

The two-dimensional reference solves the vorticity--vector-potential form of the incompressible MHD equations with a Fourier pseudo-spectral method~\cite{canuto2006}, $2/3$ de-aliasing and fourth-order integrating-factor Runge--Kutta time stepping (see e.g.~\cite{kassam2005}), with $\Delta t=2\times10^{-3}$. The runs with $N_s=256$ and $N_s=512$ give identical peak currents to the digits reported ($j_{\max}=18.26306$ at $t=0.5$ and $46.70076$ at $t=1$). The three-dimensional reference solves the velocity--magnetic-field form with the same method, with $\Delta t\le4\times10^{-3}$ for $N_s=64$ ($2\times10^{-3}$ for the cases with a guide field or a stronger field), $\Delta t\le5\times10^{-3}$ for $N_s=128$ and $\Delta t\le2.5\times10^{-3}$ for $N_s=256$, adjusted to fall on the sampling times. At $\mathrm{Re}=100$, $N_s=64$ agrees with $N_s=128$ to $10^{-8}$ in the energies. With the field multiplied by three, $N_s=64$ under-resolves the current: up to $0.9\%$ of $|\hat{\mathbf j}|^2$ lies in the three wavenumber shells closest to the de-aliasing cutoff, so that case serves only to compare variants of the lattice Boltzmann scheme with each other. At $\mathrm{Re}=1000$, $N_s=128$ is converged up to $t\approx1$; the fields of Figs.~\ref{fig:iso3d} and~\ref{fig:maps3d} come from this run, whose energies agree with those of $N_s=256$ to $2\times10^{-4}$ at $t=1$. With $N_s=256$ the solution is resolved up to $t=3$, with at most $1.1\times10^{-4}$ of $|\hat{\mathbf j}|^2$ in the three shells nearest the cut-off; that run is sampled at the output times of the lattice runs. At $\mathrm{Re}=3000$ the reference has $N_s=512$ and $\Delta t\le1.25\times10^{-3}$, and is sampled at the output times of the $N=256$ and $512$ lattice runs. No second resolution was computed there, so that its accuracy rests on the tail criterion alone: up to $t=3$ at most $8.8\times10^{-5}$ of $|\hat{\mathbf j}|^2$ lies in the three shells nearest the cut-off; its peak current is $343$ at $t=2.30$. The lattice diagnostics are compared with the spectral fields sampled on the lattice nodes, using the same finite-difference stencil.

\section{Verification of the identities}\label{app:checks}
\setcounter{table}{0}\setcounter{figure}{0}
\renewcommand{\theHfigure}{appC.\arabic{figure}}
\renewcommand{\theHtable}{appC.\arabic{table}}

The identities of Sections~\ref{sec:model} and~\ref{sec:equivalence} were checked numerically on D2Q9 and D3Q27 for random states, and the Chapman--Enskog results of Section~\ref{sec:ce} against a direct numerical evaluation of the expansion (Fig.~\ref{fig:spectra}), including a magnetic field, a body force and a separate bulk rate. The central-moment operators were implemented literally: transform, set the moments, solve the linear system. Table~\ref{tab:checks} lists the largest absolute differences, in the post-collision populations (of size $10^{-2}$--$10^{-1}$) or, for the moment identities, in the moments. The collisions of the two-dimensional RMs and CMs implementations were also transcribed and compared, state by state, with $\mathrm{PR}_{\omega,1}$ and $\mathrm{RR}_{\omega,1}$. Finally, the GPU implementation of the CHM collision on D3Q27 (Section~\ref{sec:results}), which uses esoteric-pull instead of swap streaming, was run in double precision against the RR implementation on the three-dimensional vortex ($N=32$, $\mathrm{Re}=100$, $t\le4$).

\begin{table}[h]
\centering\small
\caption{Largest absolute difference between the two sides of each identity, over 60 random states per lattice (200 for the code transcriptions). Last row: largest relative difference of the energies and $J_{\max}$ over a full three-dimensional run, for $\omega_b=\omega$ and $\omega_b=1$ alike.}
\label{tab:checks}
\begin{tabular}{@{}lcc@{}}
\toprule
identity & D2Q9 & D3Q27\\
\midrule
closed form~\eqref{eq:closed} vs.\ recursion~\eqref{eq:rr} & $1.1\times10^{-19}$ & $5.4\times10^{-20}$\\
CHM~\eqref{eq:chm} vs.\ RR, single rate & $3.3\times10^{-16}$ & $2.9\times10^{-16}$\\
CHM vs.\ RR, separate bulk rate & $2.8\times10^{-16}$ & $2.8\times10^{-16}$\\
CHM with body force vs.~\eqref{eq:force} & $2.2\times10^{-16}$ & $3.9\times10^{-16}$\\
Maxwell central moments vs.~\eqref{eq:maxwellK} & $6.9\times10^{-18}$ & $9.8\times10^{-18}$\\
CHM without Maxwell parts vs.\ RR with~\eqref{eq:comoving} & $3.3\times10^{-16}$ & $2.8\times10^{-16}$\\
monomial vs.\ Hermite central moments,~\eqref{eq:monodiff} & $4.4\times10^{-16}$ & $4.3\times10^{-16}$\\
raw Hermite, higher orders at equilibrium, vs.\ projective & $2.2\times10^{-16}$ & $5.0\times10^{-16}$\\
CMs vs.\ $\mathrm{RR}_{\omega,1}$; RMs vs.\ $\mathrm{PR}_{\omega,1}$ & $2.2\times10^{-16}$; $2.8\times10^{-16}$ & --\\
CHM vs.\ RR implementation, full run & -- & $2.6\times10^{-13}$\\
\bottomrule
\end{tabular}
\end{table}

\section{Linear stability of the coupled scheme}\label{app:lsa}
\setcounter{table}{0}\setcounter{figure}{0}
\renewcommand{\theHfigure}{appD.\arabic{figure}}
\renewcommand{\theHtable}{appD.\arabic{table}}

The RR collision~\eqref{eq:closed}, with the equilibrium~\eqref{eq:feq} and the second-order map~\eqref{eq:bulk}, and the D3Q7 collision~\eqref{eq:mag} are linearised about a uniform state $\rho=1$, $\bu=\mathbf U$, $\bb=\mathbf B_0$, at which both sets of populations are at equilibrium. The Jacobian of the 48 post-collision populations with respect to the pre-collision ones is obtained by complex-step differentiation, which is exact to round-off. For a Fourier mode $\exp(\mathrm i\mathbf k\cdot\mathbf x)$ the streaming multiplies each population by $\exp(-\mathrm i\mathbf k\cdot\bc_i)$ (by $\exp(-\mathrm i\mathbf k\cdot\mathbf e_j)$ for the field), and the eigenvalues $\lambda$ of the product give the growth $\ln|\lambda|$ per time step of each mode. 

We take the largest growth over eleven directions of $\mathbf k$ (the three lattice axes, four face diagonals, two body diagonals, $[210]$ and $[211]$) and $0<k\Delta x\le0.97\pi$, and over four directions of $\mathbf B_0$ ($[001]$, $[110]$, $[111]$ and a generic one) and, with a mean flow, two of $\mathbf U$ ($[100]$ and a generic one); at $k\Delta x=\pi$ every scheme has a neutral lattice mode with $|\lambda|=1$. The search in $k$ must be fine, because near threshold the unstable wavenumbers form a narrow band (see below). Along each direction the growth is sampled at 384 wavenumbers, 4096 for BGK whose bands are narrower still, and every local maximum that stands out from its neighbours, together with the two highest, is refined by golden-section search; on twelve states this reproduces a search over $16\,384$ wavenumbers to $2\times10^{-13}$ in the growth. The thresholds are the smallest $|\mathbf B_0|$ at which the growth exceeds $10^{-7}$ per step, found by bisection to $0.3\%$. The results do not depend on the sampling. With four times as many wavenumbers in the first pass, or with the range extended to $0.995\pi$, the thresholds of the single rate and of $\omega_b=1$ and the growth at the peak field are unchanged to the tolerance of the bisection; with 60 further directions of $\mathbf k$ and ten of $\mathbf B_0$, or, with a mean flow, 40 of $\mathbf k$, four of $\mathbf B_0$ and four of $\mathbf U$, they change by less than $1\%$, the tolerance of that test; only the threshold of the slow mode with a mean flow falls further, by $2\%$. Each unstable eigenvector is split, in the norm weighted by $1/w_i$ and $1/\psi_j$ in which the equilibrium perturbations are orthogonal to the rest, into density, longitudinal and transverse momentum, field and the non-hydrodynamic remainder; a mode is called compressive when density and longitudinal momentum carry more than $5\%$ of it. 

Table~\ref{tab:lsa} lists the thresholds at the viscosity of the $N=128$ runs at $\mathrm{Re}=3000$. The compressive mode of the regularised schemes propagates along a lattice axis and is driven by the field component across it: at the peak initial field of the runs the single rate grows by $1.2\times10^{-2}$ per step along an axis and by at most $9\times10^{-5}$ in the other eight directions. For a field across $\mathbf k$ and no mean flow its growth per step is close to $0.58\,|\mathbf B_0|-7.5\,\nu$ on $N=64$--$512$, so that the threshold is proportional to the lattice viscosity (Fig.~\ref{fig:mech}d). Both terms scale with the lattice velocity at fixed $N$ and $\mathrm{Re}$: the growth per unit time does not depend on it, $36.8$, $36.5$ and $36.5$ e-folds at twice, once and half the lattice velocity of the runs (peak field, $N=128$, no mean flow), whereas the width of the unstable band is proportional to it, $0.19$, $0.097$ and $0.049$ in $k\Delta x$. Near threshold the band is narrow, $0.0054$ at $|\mathbf B_0|=0.004$ and $0.051$ at the rms initial field, and the thresholds are those of the unbounded lattice: the wavenumbers of the periodic $N=128$ box are $0.049$ apart along an axis, and the first of its wavevectors, on the axis, enters the band at $|\mathbf B_0|\approx0.012$, which is also about what a search over a few tens of wavenumbers returns ($0.013$ with 48). The magnetic collision, D3Q7 BGK in all runs, is part of the linearised operator; its share in the thresholds is not examined here.

For the drift tests of Section~\ref{sec:res3d} the analysis is applied to the 27 fluid populations alone, without a field, about $\bu=(U_0,0,0)$ at the viscosity of those runs, for BGK and RR with the second-order and the complete equilibrium: RR with the second-order equilibrium is unstable above $U_0\approx0.095$, through a grid-scale mode along the drift at $k\Delta x\approx2.1$--$2.3$ that grows by $1.6\times10^{-3}$ per step at $U_0=0.1$ and $4.7\times10^{-2}$ at $U_0=0.2$, whereas BGK with either equilibrium and RR with the complete one are stable up to $U_0=0.2$. As a check, at $k=0.05$, $\nu=\eta=0.01$ and $|\mathbf B_0|=0.05$ across $\mathbf k$, the analysis recovers the shear and magnetic damping $\nu k^2$ and $\eta k^2$, the speed of the fast wave, $0.57947$ against $(\cs+|\mathbf B_0|^2)^{1/2}=0.57951$, and its damping, $0.00995k^2$ and $0.0620k^2$ for $\omega_b=\omega$ and $\omega_b=1$ against $(\tfrac23\nu+\tfrac12\nu_b)k^2=0.0100k^2$ and $0.0622k^2$.

\begin{table}[!htbp]
\centering\small
\caption{Field thresholds $|\mathbf B_0|$ (lattice units) of the linear instability at the viscosity of the $N=128$ runs at $\mathrm{Re}=3000$, without a mean flow and with $|\mathbf U|=0.02$, and the first mode to become unstable without a mean flow. c: compressive, propagating across the field along a lattice axis at $k\Delta x\approx2.6$; density and longitudinal momentum carry $26$--$52\%$ of it. s: slow and non-compressive, across the field at $k\Delta x\approx1$, made of transverse velocity and field; $5\%$ above threshold it grows by $2\times10^{-5}$ per step, against $1$--$3\times10^{-4}$ for c. g: $76$--$80\%$ non-hydrodynamic; across the field along a face diagonal at $k\Delta x\approx2.5$ for $\omega_b=\omega$, nearly along the field at $k\Delta x\approx2.95$ for $\omega_b=1$, a mode whose growth does not depend on the bulk rate. In brackets, the threshold of the compressive mode where it is not the first.}
\label{tab:lsa}
\begin{tabular}{@{}llccc@{}}
\toprule
collision & $\omega_b$ & no mean flow & $|\mathbf U|=0.02$ & first unstable mode\\
\midrule
BGK & $\omega$ & $0.0022$ & $0.0025$ & g\\
BGK & $1$ & $0.0028$ & $0.0034$ & g\\
PR & $\omega$ & $0.0037$ & $0.0036$ & c\\
PR & $1$ & $0.022$ & $0.023$ & s\\
RR & $\omega$ & $0.0037$ & $0.0036$ & c\\
RR & $1.9$ & $0.012$ & $0.012$ & c\\
RR & $1.8$ & $0.017$ & $0.016$ & c\\
RR & $1.5$ & $0.022$ ($0.027$) & $0.022$ & s\\
RR & $1$ & $0.022$ ($0.038$) & $0.023$ ($0.037$) & s\\
\bottomrule
\end{tabular}
\end{table}

\section{The Lorentz correction in practice}\label{app:lorentz}
\setcounter{table}{0}\setcounter{figure}{0}
\renewcommand{\theHfigure}{appE.\arabic{figure}}
\renewcommand{\theHtable}{appE.\arabic{table}}

With the BGK collision the populations carry the actual non-equilibrium. Its third-order Hermite coefficients can be compared with two predictions from the measured $\bm a^{(2)}_1$: the hydrodynamic recursion~\eqref{eq:recursion}, and the recursion with the correction~\eqref{eq:delta} at leading order. Figure~\ref{fig:ce} shows the relative residual, i.e.\ the relative error of each prediction in the norm of the full tensor summed over the nodes, for four configurations: no field, the standard field, the standard field plus a uniform guide field $3\hat{\mathbf z}$, and three times the standard field.
\begin{itemize}
\item \emph{No field.} The recursion is consistent with the first-order solution; the residual of $3.7\%$ at $N=64$ is that of the first-order solution itself.
\item \emph{With a field.} Once the start-up transient has passed ($t\ge0.08$ on $N=64$ and $128$, $t\ge0.3$ on $N=32$), the relative error of the hydrodynamic recursion is $78$--$96\%$ up to $t=0.5$, on every grid. At $t\approx0.2$ the corrected prediction is as accurate as the recursion is without a field ($2$--$4\%$); later its residual grows as the sheets steepen, by $t=0.5$ to $5\%$ for the standard field and $23\%$ for the strongest one on $N=64$.
\item \emph{Grid convergence.} For the strongest field the corrected residual converges with the grid: $63\%$, $17\%$ and $6.1\%$ for $N=32$, $64$ and $128$ at $t\approx0.3$, against $88$--$94\%$ for the recursion.
\item \emph{$\mathrm{Ma}^2$ terms.} Adding them changes the residuals by less than $10^{-3}$.
\end{itemize}

\begin{figure}[!htbp]
\centering
\begin{tikzpicture}
\begin{groupplot}[group style={group size=2 by 1, horizontal sep=1.7cm}, width=0.47\linewidth, height=5.8cm, grid=major, grid style={gray!20},
  legend style={draw=none, fill=white, fill opacity=0.85, text opacity=1, font=\scriptsize}, legend cell align=left]
\nextgroupplot[xlabel={$t$}, ylabel={relative residual of $\bm a_1^{(3)}$}, title={(a) $N=64$}, ymode=log, xmin=0.06, xmax=0.52, ymin=8e-3, ymax=1.5,
  legend to name=leg:ce, legend columns=4, legend style={/tikz/every even column/.append style={column sep=0.3cm}}]
\addplot[cBGK, thick] table[x=t, y=hydro] {data/ce_b0_N64.dat};
\addplot[cRR, thick, densely dashed] table[x=t, y=hydro] {data/ce_std_N64.dat};
\addplot[cCM, thick, densely dashed] table[x=t, y=hydro] {data/ce_guide3_N64.dat};
\addplot[cRM, thick, densely dashed] table[x=t, y=hydro] {data/ce_bs3_N64.dat};
\addplot[cRR, thick] table[x=t, y=corr1] {data/ce_std_N64.dat};
\addplot[cCM, thick] table[x=t, y=corr1] {data/ce_guide3_N64.dat};
\addplot[cRM, thick] table[x=t, y=corr1] {data/ce_bs3_N64.dat};
\legend{$\mathbf b=\mathbf 0$, standard: recursion, guide field: recursion, $3\mathbf b$: recursion, standard: corrected, guide field: corrected, $3\mathbf b$: corrected}
\nextgroupplot[xmode=log, ymode=log, log basis x=2, xtick={32,64,128}, xticklabels={32,64,128}, xmin=26, xmax=158, ymin=1.5e-3, ymax=1.5,
  xlabel={$N$}, title={(b) field $3\mathbf b$}, legend pos=south west, legend style={font=\scriptsize}]
\addplot[cRM, thick, densely dashed, mark=o] table[x=N, y=hydro03] {data/ce_conv.dat};
\addplot[cRM, thick, mark=*] table[x=N, y=corr03] {data/ce_conv.dat};
\addplot[cRM!50, thick, mark=square*] table[x=N, y=corr02] {data/ce_conv.dat};
\addplot[black, dashed, domain=32:128, samples=2] {0.63*(32/x)^2};
\legend{{recursion, $t\approx0.3$}, {corrected, $t\approx0.3$}, {corrected, $t\approx0.2$}, slope $-2$}
\end{groupplot}
\node[anchor=south] at ($(group c1r1.north)!0.5!(group c2r1.north)+(0,0.95cm)$) {\pgfplotslegendfromname{leg:ce}};
\end{tikzpicture}
\caption{Relative residual of the third-order non-equilibrium Hermite coefficients of the BGK populations, predicted from the measured second-order one by the hydrodynamic recursion (dashed) and with the Lorentz correction (solid); three-dimensional vortex, $\mathrm{Re}=100$. (a) $N=64$; (b) convergence for the strongest field.}
\label{fig:ce}
\end{figure}

\begin{table}[!htbp]
\centering\footnotesize
\caption{Effect of the Lorentz correction on RR at $N=64$, $\mathrm{Re}=100$: maximum relative errors (\%) of $E_{\rm k}$ / $E_{\rm m}$ / $J_{\max}$ against the spectral solution for $0.25\le t\le4$ and, in the last column, maximum relative difference (\%) between the two runs over the same interval; last two rows: divergence time at $\mathrm{Re}=3000$, $N=64$ (samples every $0.017$). Level 1 is~\eqref{eq:df}; level 2, which adds the $O(\mathrm{Ma}^2)$ terms, changes the energies by less than $5\times10^{-7}$, $J_{\max}$ by less than $6\times10^{-6}$ and the divergence times not at all, and is not shown. For the field $\times3$ the spectral reference is under-resolved (\ref{app:spectral}).}
\label{tab:lorentz}
\begin{tabular}{@{}lccc@{}}
\toprule
 & hydrodynamic recursion & Lorentz-corrected (level 1) & difference of the two runs\\
\midrule
standard field & 0.196 / 0.321 / 2.66 & 0.193 / 0.324 / 2.62 & 0.035 / 0.017 / 0.11\\
guide field $3\hat{\mathbf z}$ & 0.455 / 0.105 / 2.65 & 0.471 / 0.105 / 2.70 & 0.024 / 0.0025 / 0.10\\
field $\times3$ & 0.609 / 0.493 / 5.46 & 0.601 / 0.530 / 5.33 & 0.096 / 0.047 / 0.46\\
\midrule
$\mathrm{Re}=3000$, standard & $t=1.65$ & $t=1.60$ & \\
$\mathrm{Re}=3000$, guide field & $t=1.57$ & $t=1.36$ & \\
\bottomrule
\end{tabular}
\end{table}

The macroscopic effect of the correction is below the resolution error (Table~\ref{tab:lorentz}): the runs with and without it differ from each other by $5$ to $41$ times less than they differ from the spectral solution. The fields differ as little: on $N=64$ with the standard field, the relative $L^2$ difference between the two runs at $t=1$, $2$ and $3$ is $0.07$--$0.08\%$ in $|\mathbf j|$, $0.1$--$0.3\%$ in the vorticity magnitude, which measures the velocity gradients and hence the viscous stress, $0.08$--$0.09\%$ in $\bu$ and $0.03$--$0.06\%$ in $\bb$. The difference behaves as a truncation error does, in agreement with the order estimate of Section~\ref{sec:ce}: it falls by factors of $5.2$, $7.3$ and $3.5$ in $E_{\rm k}$, $E_{\rm m}$ and $J_{\max}$ between $N=32$ and $64$, and of $3.2$, $4.6$ and $4.5$ between $N=64$ and $128$, faster than the errors themselves ($4.1$, $3.8$ and $2.8$, then $2.6$, $3.3$ and $4.0$). Halving the lattice velocity on the same grid reduces it by factors of $1.1$ to $1.5$ only. The correction does not improve robustness either: at $\mathrm{Re}=3000$ and $N=64$, a grid far too coarse to resolve the dissipation scale, the corrected scheme diverges somewhat earlier than the hydrodynamic one.

\section*{Declaration of generative AI and AI-assisted technologies in the manuscript preparation proces}
During the preparation of this manuscript, the authors used Claude Fable 5.2 (Anthropic) to assist with language and figure editing. The authors critically reviewed and edited all AI-assisted content and take full responsibility for the accuracy, originality, and integrity of the manuscript.

\section*{Declaration of competing interest}
The author declares no competing interests.

\bibliographystyle{elsarticle-num}
\bibliography{mhd_references}

\end{document}